\documentclass[%
 reprint,
superscriptaddress,
 amsmath,amssymb,
prb,
]{revtex4-2}
\usepackage[none]{hyphenat}
\usepackage{graphicx}
\usepackage{dcolumn}
\usepackage{bm}
\usepackage{hyperref}

\hypersetup{
  colorlinks   = true, 
  urlcolor     = blue, 
  linkcolor    = blue, 
  citecolor   = blue 
}
\usepackage[version=4]{mhchem}
\usepackage{amsmath}
\usepackage{tikz}
\usetikzlibrary{shapes.geometric, arrows, calc, backgrounds}

\usepackage{natbib}

\usepackage{tikz}
\usetikzlibrary{arrows.meta,
                chains,
                decorations.pathreplacing,calligraphy,
                positioning,
                quotes,
                shapes.geometric
                }

\tikzstyle{startstop} = [rectangle, rounded corners, 
minimum width=3cm, 
minimum height=1cm,
text centered, 
draw=black, 
fill=red!30]

\tikzstyle{io} = [trapezium, 
trapezium stretches=true, 
trapezium left angle=70, 
trapezium right angle=110, 
minimum width=3cm, 
minimum height=1cm, text centered, 
draw=black, fill=blue!30]

\tikzstyle{process} = [rectangle, 
minimum width=3cm, 
minimum height=1cm, 
text centered, 
text width=3cm, 
draw=black, 
fill=orange!30]

\tikzstyle{decision} = [diamond, 
minimum width=3cm, 
minimum height=1cm, 
text centered, 
draw=black, 
fill=green!30]
\tikzstyle{arrow} = [thick,->,>=stealth]

\usepackage{enumitem}
\usepackage[caption=false]{subfig}

\usepackage{relsize}

\begin{document}
\setcitestyle{super}

\preprint{APS/123-QED}

\title{
Autonomous data extraction from peer reviewed literature for training machine learning models of oxidation potentials}

\author{Siwoo Lee}
\affiliation{Department of Chemistry, University of Toronto, St. George campus, Toronto, ON, Canada}

\author{Stefan Heinen}
\affiliation{Vector Institute for Artificial Intelligence, Toronto, ON, M5S 1M1, Canada}

\author{Danish Khan}
\affiliation{Department of Chemistry, University of Toronto, St. George campus, Toronto, ON, Canada}
\affiliation{Vector Institute for Artificial Intelligence, Toronto, ON, M5S 1M1, Canada}

\author{O. Anatole von Lilienfeld}%
\email{anatole.vonlilienfeld@utoronto.ca}
\affiliation{Department of Chemistry, University of Toronto, St. George campus, Toronto, ON, Canada}
\affiliation{Vector Institute for Artificial Intelligence, Toronto, ON, M5S 1M1, Canada}
\affiliation{Acceleration Consortium, University of Toronto. 80 St George St, Toronto, ON M5S 3H6}
\affiliation{Department of Materials Science and Engineering, University of Toronto, St. George campus, Toronto, ON, Canada}
\affiliation{Department of Physics, University of Toronto, St. George campus, Toronto, ON, Canada}
\affiliation{Machine Learning Group, Technische Universität Berlin and Berlin Institute for the Foundations of Learning and Data, Berlin, Germany}

\date{\today}

\begin{abstract}
We present an automated data-collection pipeline involving a convolutional neural network and a large language model 
to extract user-specified tabular data from peer-reviewed literature.
The pipeline is applied to 74 reports published between 1957 and 2014 with experimentally-measured oxidation potentials for 592 organic molecules (-0.75--3.58 V). 
After data curation (solvents, reference electrodes, and missed data points), we trained multiple supervised machine learning models reaching prediction errors similar to
experimental uncertainty ($\sim$0.2 V).
For experimental measurements of identical molecules reported in multiple studies, we identified the most likely value based on out-of-sample machine learning predictions.
Using the trained machine learning models, we then estimated oxidation potentials of $\sim$132k small organic molecules from the QM9 data set, with predicted values spanning 0.21--3.46 V.
Analysis of the QM9 predictions in terms of plausible descriptor-property trends suggests that aliphaticity increases the oxidation potential of an organic molecule 
on average from $\sim$1.5 V to $\sim$2 V, while an increase in number of heavy atoms lowers it systematically. 
The pipeline introduced offers significant reductions in human labor otherwise required for conventional manual data collection of experimental results, and exemplifies how to accelerate scientific research through automation.

\end{abstract}

\maketitle


\section{\label{sec:introduction}Introduction}

The accessibility and utilization of literature data through systematic reviews and meta-analyses are of significant importance across all scientific disciplines to rigorously assess the wealth of information contained in multiple studies and compile them in large-scale data sets \cite{What_is_the_difference_between_a_systematic_review_and_a_meta_analysis, introduction_to_systematic_review_and_meta_analysis, systematic_review_brief_overview_of_methods_limitations_and_resources, development_testing_and_use_of_data_extraction_forms_in_systematic_reviews}.
However, reproducibility concerns as well as the rapid growth in the number of scientific publications \cite{growth_rates_of_modern_science, the_rate_of_growth_in_scientific_publication_and_the_decline_in_coverage_provided_by_Science_Citation_Index} poses significant limitations on efficiently reading, understanding, and extracting the enormous volume of ever growing information.
The development of automated retrieval of pertinent information\cite{challenges_and_advances_in_information_extraction_from_scientific_literature} could address the challenge of training meaningful machine learning (ML) models that require sufficiently large scientific data sets \cite{machine_learning_trends_perspectives_and_prospects, training_set_size_requirements_for_the_classification_of_a_specific_class}.
In particular, tabular data in literature sources holds immense importance in scientific research as they organize a large body of information in an easily-readable fashion. 
Thus, the efficient extraction of tabular information would greatly streamline data collection from a large number of studies. 
Yet, upon examining different reference sources, it is evident that tables are presented in a variety of layouts, visual appearances, and encoding formats (eg. HTML, PDF, JPG), which poses a significant hurdle in the automated detection of tables in the literature\cite{current_status_and_performance_analysis_of_table_recognition_in_document_images_with_deep_neural_networks}. 
However, recent advances in algorithmic designs and computing capabilities have seen the development of convolutional neural network (CNN) models, such as TableNet\cite{current_status_and_performance_analysis_of_table_recognition_in_document_images_with_deep_neural_networks, tablext_a_combined_neural_network_and_heuristic_based_table_extractor, tablenet_deep_learning_model_for_end_to_end_table_detection_and_tabular_data_extraction_from_scanned_document_images}, that are trained to locate tables in document pages displayed as images and are capable of reaching state-of-the-art performances on the ICDAR 2013 table competition data set \cite{ICDAR_2013_table_competition}. 
A secondary challenge that follows table detection using CNN models is the accurate extraction of text from images, a task known as optical character recognition (OCR) \cite{a_survey_on_optical_character_recognition_system}. 
Google's \texttt{Tesseract-OCR} engine \cite{an_overview_of_the_tesseract_OCR_engine, history_of_the_tesseract_ocr_engine_what_worked_and_what_did_not} and various ML and deep neural network (DNN) models have been demonstrated to successfully convert images of typed, handwritten, or printed text into machine-encoded text with low character-level substitution rates and word-level error-rates \cite{history_of_the_tesseract_ocr_engine_what_worked_and_what_did_not, handwritten_optical_character_recognition}. 
Then, a third, and closely-related problem relevant to scientific research is the ability of these models to extract specific text. 
This presents a significant challenge due to the need for semantic understanding, especially as documents may display several tables containing different types of data with irrelevant accompanying information \cite{tabular_data_semantic_understanding}.
The recent development of large language models (LLMs) presents a promising solution to the challenge of semantic understanding as they can leverage their extensive training on large volumes of text to recognize and interpret the meaning of specified text \cite{a_survey_of_large_language_models}.
Indeed, LLMs have already seen widespread usage for a variety of scientific purposes \cite{a_bibliometric_review_of_large_language_models_research_from_2017_to_2023}. 
For instance, in chemistry, LLMs have been utilized to generate code, learn complex molecular distributions, aid in materials and drug design, and to extract chemical information from scientific documents \cite{language_models_can_learn_complex_molecular_distributions, chemical_language_models_for_de_novo_drug_design, natural_language_processing_models_that_automate_programming_will_transform_chemistry_research_and_teaching,is_GPT3_all_you_need_for_low_data_discovery_in_chemistry, material_transformers_deep_learning_language_models_for_generative_materials_design, chemdataextractor}. 
Generative pre-trained transformers (eg. GPT-2, GPT-3.5, GPT-4) models developed by OpenAI present particularly exciting applications for research in chemistry and other scientific disciplines for their human-like semantic understanding and their ability to generate human-like text when presented with a prompt \cite{introducing_chatgpt, Gpts_are_gpts, openai2023gpt, GPT4_Vs_GPT3_5_a_concise_showdown, chatgpt_and_other_large_language_models_are_double_edged_swords}.\\

In this work, an automated data-collection pipeline is introduced that accurately locates tables and extracts text from literature sources using the CNN TableNet, and the LLM GPT-3.5, respectively. 
We demonstrate its usefulness by building a chemically-diverse data set of experimentally-measured oxidation potentials (measured in acetonitrile solvent vs standard calomel electrode, SCE) of organic molecules from peer-reviewed literature.
Oxidation potentials are important electrochemical stability and reactivity descriptors; modeling them with efficient machine learning and high predictive power could crucially accelerate the computational design and discovery of superior functional materials, such as batteries, supercapacitors, electrolytes, and electrocatalysts for applications in fuel cells and renewable energy conversion \cite{organic_electrolytes_for_redox_flow_batteries, new_insights_into_phenazine_based_RFBs, organic_flow_batteries_recent_progress_and_perspectives, recent_advancements_in_rational_design_of_nonaqueous_RFBs}. 
Based on the experimental data extracted using our pipeline, we have trained multiple supervised ML models that reach experimental uncertainty, and that can be used to identify less/more likely values among conflicting data entries.
The generalizability of the ML models is used to predict and analyze the oxidation potential distribution in $\sim$132k organic molecules coming from the QM9 data set \cite{QM9_database}.
Previous ML studies of redox potentials of organic molecules were limited to small data sets based on simulated values which typically encode severe approximations making it difficult to draw direct conclusions relevant for experimental decision making.\cite{computational_electrochemistry, DFT_computation_of_redox_potentials_in_solution, Anthraquinone_redox_chemistry_DFT, ML_corrections_to_DFT_calculated_redox_potentials, implicit_vs_explicit_solvent_models, implicit_solvation_model_review, machine_learning_properties_of_electrolyte_additives, predicting_redox_potentials_of_phenazine_derivatives, automated_workflow_for_computation_of_redox_potentials, molecular_structure_redox_potential_relationship_for_organic_electrode_materials}.

\section{\label{sec:Methodology}Methodology}

\subsection{\label{sec:pipeline} From Literature to Data Set}

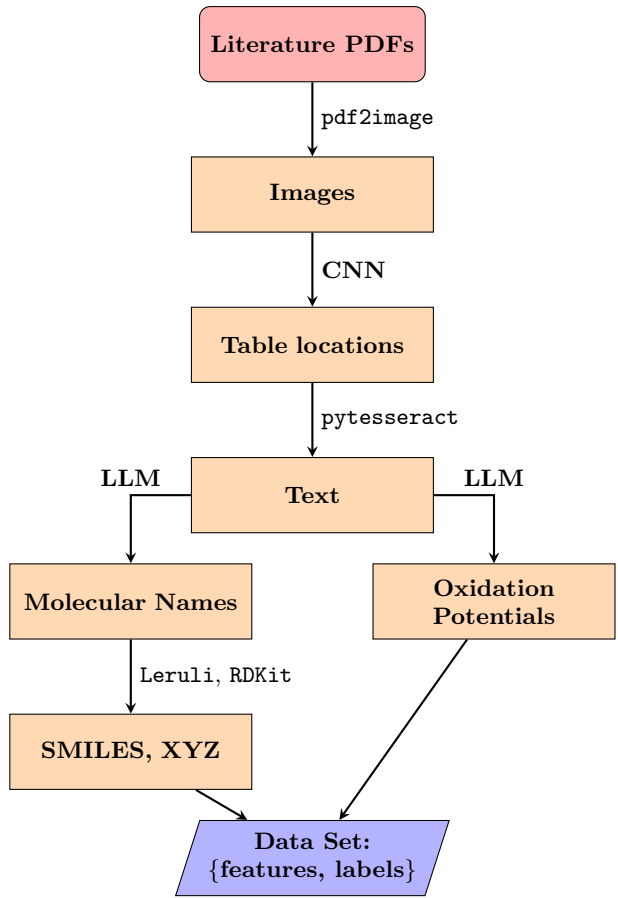
\begin{figure}
  \centering
  \begin{tikzpicture}[node distance=2cm]
    \node (start) [startstop] {\textbf{Literature PDFs}};
    \node (pro0) [process, below of=start] {\textbf{Images}};
    \node (pro1) [process, below of=pro0] {\textbf{Table locations}};
    \node (pro2) [process, below of=pro1] {\textbf{Text}};
    \node (pro3a) [process, below left of=pro2, xshift=-1cm] {\textbf{Molecular Names}};
    \node (pro3b) [process, below right of=pro2, xshift=1cm] {\textbf{Oxidation Potentials}};
    \node (pro4) [process, below of=pro3a] {\textbf{SMILES, XYZ}};
    \node (stop) [io, below right of=pro4, xshift =1cm, align=center] {\textbf{Data Set:} \\  \{\textbf{features, labels}\}};

    \draw [arrow] (start) -- (pro0) node[midway, right] {\texttt{pdf2image}};
    \draw [arrow] (pro0) -- (pro1) node[midway, right] {\textbf{CNN}};
    \draw [arrow] (pro1) -- (pro2) node[midway, right] {\texttt{pytesseract}};
    \draw [arrow] (pro2.west) -| (pro3a) node[midway, above] {\textbf{LLM}};
    \draw [arrow] (pro2.east) -| (pro3b) node[midway, above] {\textbf{LLM}};
    \draw [arrow] (pro3a) -- (pro4) node[midway, right] {\texttt{Leruli}, \texttt{RDKit}};
    \draw [arrow] (pro3b) -- (stop) node[midway, above]{};
    \draw [arrow] (pro4) -- (stop) node[midway, above]{};

  \end{tikzpicture}

    \caption{A flowchart representation of the automated data acquisition pipeline for extracting experimentally-measured oxidation potentials reported in literature. Pages displaying tables in 74 reference sources (PDFs) are converted to images (JPGs) and inputted into a convolutional neural network, TableNet, trained to locate tables in images and output images cropped around tables. Text contained in the outputted images, extracted using \texttt{pytesseract}\cite{Pytesseract} are then fed to GPT-3.5 with a prompt to extract the names of molecules and their oxidation potentials. Used \texttt{Python} packages for each step are shown beside the arrows.}
  \label{fig: pipeline flowchart}
  
\end{figure}

The first component of the automated tabular data extraction pipeline (\autoref{fig: pipeline flowchart}) after the collection of literature sources is the detection and localization of tables (\textbf{CNN} step of \autoref{fig: pipeline flowchart}). 
This is accomplished by using TableNet with a DenseNet-121 encoder architecture (8,220,550 trainable parameters; 461,504 non-trainable parameters) with dropout (0.6) \cite{densely_connected_convolutional_networks} (see \textit{Section J} of Supplementary Information for the \texttt{Python} implementation used in this work and \textcite{tablenet_deep_learning_model_for_end_to_end_table_detection_and_tabular_data_extraction_from_scanned_document_images} for further details about the architecture). 
This model was trained for 35 epochs on 495 scanned RGB images (816 $\times$ 1056 pixels) of document pages containing tables with English text compiled in the Marmot data set (80/20 train/test random split) with labelled coordinates of the rectangular table regions in each image \cite{dataset_ground_truth_and_performance_metrics_for_table_detection_evaluaiton}. 
The model learns these coordinates such that it can output cropped images of the documents containing just the detected tables.\\

The generalization capabilities of the CNN were then assessed by its ability to locate tables in 74 literature sources (published 1957-2014), saved as PDFs, that reported the experimentally-measured oxidation potentials of organic molecules (see \textit{Bibliography} of Supplementary Information for the used references). 
\texttt{pdf2image}\cite{pdf2image} was used to convert the PDF pages to JPGs (816 $\times$ 1056 pixels), which were inputted into the CNN.
The text contained in the outputted cropped images were extracted using \texttt{pytesseract}\cite{Pytesseract}, the Python wrapper for \texttt{Tesseract-OCR} (\texttt{pytesseract} step of \autoref{fig: pipeline flowchart}). 
The blocks of text were each individually forwarded into the GPT-3.5 API once to screen for data of oxidation potentials with the following prompt (\textbf{LLM} steps of \autoref{fig: pipeline flowchart}):\\\\
\textit{Does this following piece of text contain one or more tables of oxidation potentials of organic molecules? If it does, give the code for a neatly-displayed Panda DataFrame explicitly listing only the molecules and their corresponding oxidation potentials. Ensure to list all molecules. Also, if stated, report the reference electrode and the solvent the measurements occurred in.}\\

If GPT-3.5 was able to successfully output the name of molecules, their oxidation potentials, and the reference electrode and solvent used in the experimental measurements, the master data set was compiled by including only neutrally-charged samples measured in acetonitrile to account for typical electrochemical measurement conditions in the laboratory \cite{beginners_guide_to_CV}. 
For samples labelled by their full names, the \texttt{Leruli} API\cite{LERULI} was used to convert the names to their canonical SMILES\cite{SMILES}, followed by the use of \texttt{RDKit}\cite{RDKit} to produce XYZ files from the SMILES (\texttt{Leruli}, \texttt{RDKit} steps of \autoref{fig: pipeline flowchart}). 
The XYZ files were inputted into the extended tight binding (\texttt{XTB}) API\cite{XTB_METHODS, XTB_PYTHON} to produce (implicit solvation) optimized geometries in acetonitrile.
\texttt{XTB} also produced 17 calculated values for each molecule, including their HOMO-LUMO gaps and solvation free energies in acetonitrile. 
The oxidation potentials of molecules measured in multiple studies were taken as the mean value. 
Measurements referenced against non-SCE electrodes were converted to be referenced against SCE as according to handbooks on the standard potentials of reference electrodes \cite{electrochemistry_in_nonaqueous_solutions, handbook_of_reference_electrodes}. 
The data set was supplemented as necessary by human labor for samples that the pipeline missed or incorrectly reported, as well as for cases in which the reference electrodes and solvents used in the experimental measurements could not be determined from the text contained in the tables.\\

\subsection{\label{sec:ML algorithms}eXtreme Gradient Boosting and Kernel Ridge Regression}

XGBoost Regression (eXtreme Gradient Boosting Regression: XGBR) was selected as a candidate ML algorithm due to its exceptional performance and versatility in handling various regression tasks due to gradient boosting and optimized tree-based ensemble learning algorithms \cite{XGBoost}.\\

Kernel ridge regression (KRR) was also tested as it is a popular algorithm for ML in quantum chemistry due to its ease of hyperparameter tuning, in addition to its excellent ability to capture non-linear relationships using kernel functions and its efficiently handing of high-dimensional data \cite{kernel_ridge_regression_in_quantum_chemistry, Fundamentals_of_quantum_machine_learning}. 
It accomplishes this using kernel functions, which in this work are selected to be Laplacian kernels of the form 

\begin{equation}
    K(\textbf{A}_{i}, \textbf{B}_{j}) = \exp{\left(-\dfrac{ || \textbf{A}_{i} - \textbf{B}_{j} ||_{1}}{\sigma} \right)}
\end{equation}
where $\textbf{A}_{i}, \textbf{B}_{j}$ denote the representation vectors of molecules $i, j$\cite{QML_library, sklearn_KRR}.\\

Bayesian optimization implemented with \texttt{hyperopt}\cite{HYPEROPT} was used for hyperparameter-tuning both algorithm types, with hyperparameters selected as those that returned the lowest mean absolute error, MAE, on four-fold cross-validation on the training set (80/20 train/test random split). \\

\subsection{\label{sec:reps} Physics-Based Structural Representations}

Four XGBR models were developed in this work with the following input features: ACSF\cite{atom_centered_symmetry_functions}; ACSF, XTB values; ACSF, MORDRED \cite{MORDRED}; ACSF, XTB values, MORDRED. 
MORDRED is a popular two- and three-dimensional molecular descriptor-calculation software in cheminformatics and is used, in this work, to generate three-dimensional descriptors from MOL files produced from the \texttt{XTB}-geometry-optimized XYZ coordinates. 
Three KRR models were also developed with input features of ACSF, SOAP\cite{smooth_overlap_of_atomic_positions}, and SLATM\cite{SLATM}.\\

The XYZ files were used to produce three popular physics-inspired structural representations\cite{Physics_inspired_representations} of atomic and molecular environments: atom-centered symmetry functions (ACSF), smooth overlap of atomic positions (SOAP)\cite{smooth_overlap_of_atomic_positions}, and Spectrum of London and Axilrod-Teller-Muto potentials (SLATM) \cite{SLATM}. 
These representations were used to predict the oxidation potentials of organic molecules using three KRR models.\\

ACSFs are local descriptors that express a molecule's total energy as a sum of atomic energies by constructing many-body symmetry functions, composed of radial and angular parts, for all atoms within a specified cutoff radius as given by a cutoff function\cite{atom_centered_symmetry_functions}.
This work uses radial symmetry functions of

\begin{equation}
    G_{i}^{2} = {\mathlarger{\sum}}_{j} \exp{\left( -\eta (R_{ij} - R_{s})^{2} \right) } \cdot f_{c}(R_{ij})
\end{equation}
where $\eta$ defines the width of the Gaussian function and $R_{s}$ shifts the Gaussian functions by a certain radial distance\cite{atom_centered_symmetry_functions}.
This work uses angular symmetry functions of 

\begin{eqnarray}
    G_{i}^{4} = 2^{1 - \zeta} \mathlarger{\sum}_{j,k \neq i}^{all} \left(1 + \lambda \cos\theta_{ijk} \right)^{\zeta}\nonumber\\ 
    \cdot \exp{\left( -\eta (R_{ij}^2 + R_{ik}^2 + R_{jk}^2) \right) }\nonumber\\
    \cdot f_{c}(R_{ij}) \cdot f_{c}(R_{ik}) \cdot f_{c}(R_{jk})
\end{eqnarray}
where $\zeta$ defines the angular resolution of the symmetry functions and $\lambda$ shifts the maxima of the cosine functions between 0 and $\pi$ radians\cite{atom_centered_symmetry_functions}.
The ACSF representations are generated using the DScribe library\cite{DScribe} with $R_{c} =$ 9.0 \AA, 6 pairs of $\eta, R_{s}$ parameters for the $G^{2}$ radial functions, and 6 triplets of $\eta, \zeta, \lambda$ parameters for the $G^{4}$ angular functions.\\

SOAP descriptors represent local atomic environments where each is described by a single power spectrum of the form

\begin{equation}
    p(\textbf{r})_{n, n', l}^{a_{1} a_{2}} = \pi \sqrt{\dfrac{8}{2l + 1}} \mathlarger{\sum}_{m} c_{n, l, m}^{a_{1}}(\textbf{r})^{\dagger} c_{n, l, m}^{a_{2}}(\textbf{r})
\end{equation}
where $a_{1}, a_{2}$ index different atoms\cite{smooth_overlap_of_atomic_positions, DScribe}.
DScribe was again used to generate SOAP representations in this work, with parameters selected as $n_{\text{max}} = 6$ (maximum number of radial basis functions), $l_{\text{max}} = 6$ (maximum degree of spherical harmonics), $\sigma = 0.1$, and spherical Gaussian-type orbitals for the radial basis functions, $g_{n}$.\\

SLATM returns a global representation of the charge density of a given system by concatenating different many-body potential spectra composed of one, two, and three-body terms representing the atomic nuclear charges, London potentials, and Axilrod-Teller-Muto van der Waals potentials, respectively\cite{SLATM}.
In this work, SLATM representations were generated using the QML-code library \cite{QML_library}.\\

The best-performing ML model on the test set was then used to screen the oxidation potentials of $\sim$132k molecules listed in the QM9 database\cite{QM9_database, QM9_database2}, which reports the geometries of $\sim$134k stable small organic molecules with up to 9 heavy (non-hydrogen) atoms (C, N, O, F) computed at the B3LYP/6-31G(2df,p) level of quantum chemistry\cite{Basis, Functional, Functional2}. 
The molecules in QM9 thus lie within the domain of the extracted data set by chemical composition and is suitable for estimations of oxidation potentials by the developed ML model based on interpolations.

\section{\label{sec:results and discussion}Results and Discussion}

\subsection{\label{sec:Collected Data Set}Extracting Data}


The performance of the automated data collection pipeline in accurately identifying tables containing oxidation potentials and extracting their values was verified via human labor. 
In the 74 reference sources, one human count returned a total of 182 tables, containing a variety of information such as oxidation potentials, spectroscopic data, product yields, and reaction kinetics. 
Of these, the CNN failed to locate 19 tables, a 10 \% error which is comparable to that associated with some top-performing table detection models \cite{current_status_and_performance_analysis_of_table_recognition_in_document_images_with_deep_neural_networks}(\autoref{fig: CNN table accuracy}) (see \textit{Section I} of Supplementary Information for an example output from the CNN).\\



\begin{figure}[h]
\resizebox{\columnwidth}{!}{\includegraphics{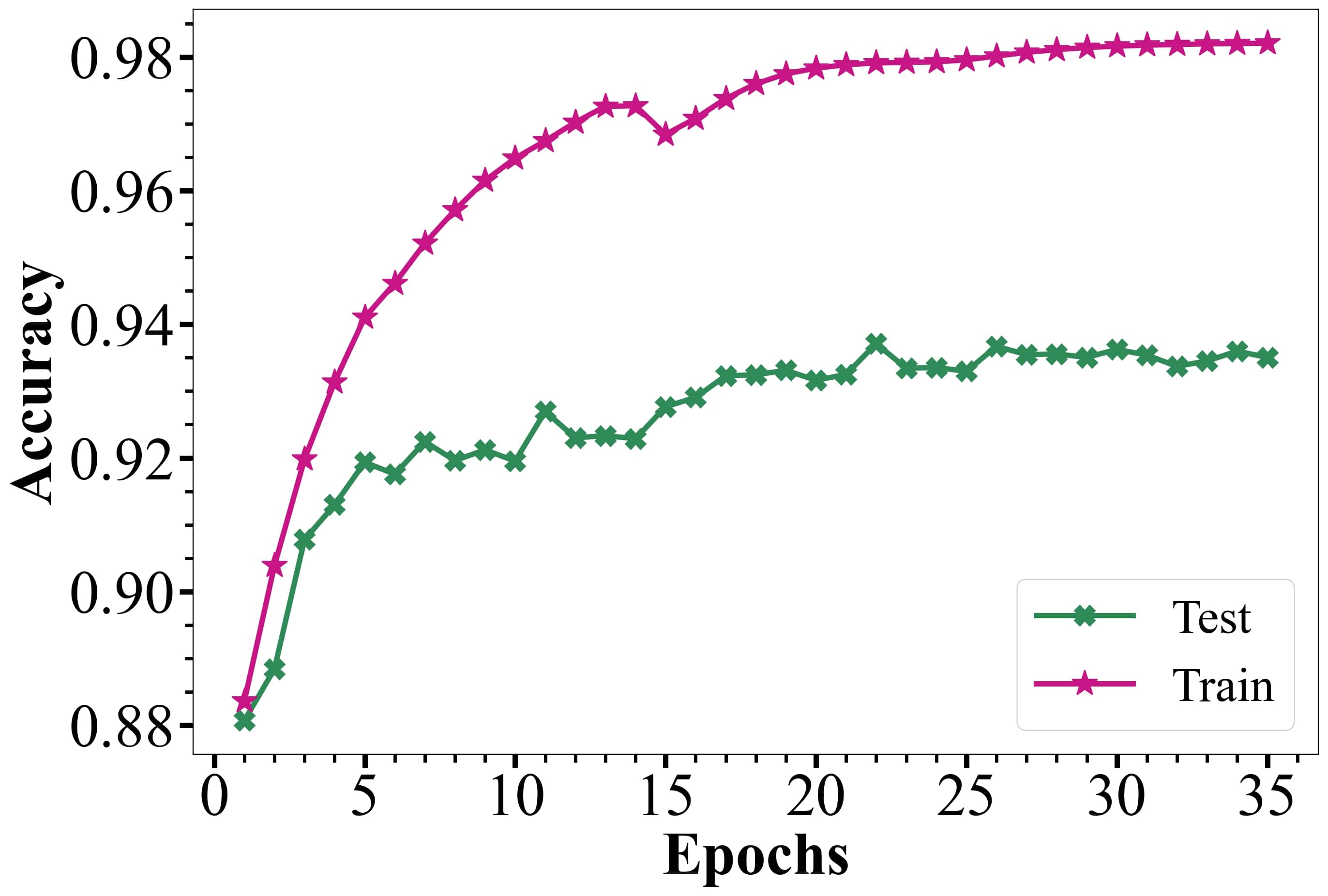}}
\caption{\label{fig: CNN table accuracy}Performance of CNN on training and testing (80/20 random split) of the Marmot data set\cite{dataset_ground_truth_and_performance_metrics_for_table_detection_evaluaiton} evaluated as accuracy of detecting tables (percent overlap of predicted table location area with actual area), vs. number of training epochs.}
\end{figure}

The extracted text from the table images outputted from the CNN were then forwarded into GPT-3.5 to screen for measurements of oxidation potentials. 
One human count returned a total of 1715 measurements. 
GPT-3.5 failed to accurately report the oxidation potentials of 445 samples (26 \% error) (see \textit{Section I} of Supplementary Information for an example output from GPT-3.5). 
However, 262 instances of these were due to the molecular samples being labelled with bond-line structures, numbers, or by their substituent groups. 
171 samples were simply missed by GPT-3.5, and 12 samples had incorrectly reported oxidation potentials. 
Therefore, only considering samples that were not detected or were incorrectly reported, GPT-3.5 yields an error rate of 13\%. 
The data extraction performance may be improved by including optical chemical structure recognition tools to screen for molecular names and SMILES of compounds represented as bond-line structures\cite{decimer, decimer1, decimer_segmentation}.\\

The compiled data set includes 592 unique molecules with a range of oxidation potentials of -0.75--3.58 V, with a mean value of 1.32 V (\autoref{fig: Collected data set oxidation potentials distribution}).
See \textit{Section A} of Supplementary Information for the table listing the oxidation potentials of all molecules.
On average the molecules have a molar mass of 184 g/mol (28--680 g/mol), 26 atoms (5--86 atoms), and 13 heavy atoms (2--46 heavy atoms) (see \textit{Section B} of Supplementary Information for distribution plots of these parameters).
Out of these 592 molecules, for 155 molecules there are multiple entries in the literature; their deviations are shown in \autoref{fig: Collected data set oxidation potentials min max ranges}.
\\
\begin{figure}[h]
\resizebox{\columnwidth}{!}{\includegraphics{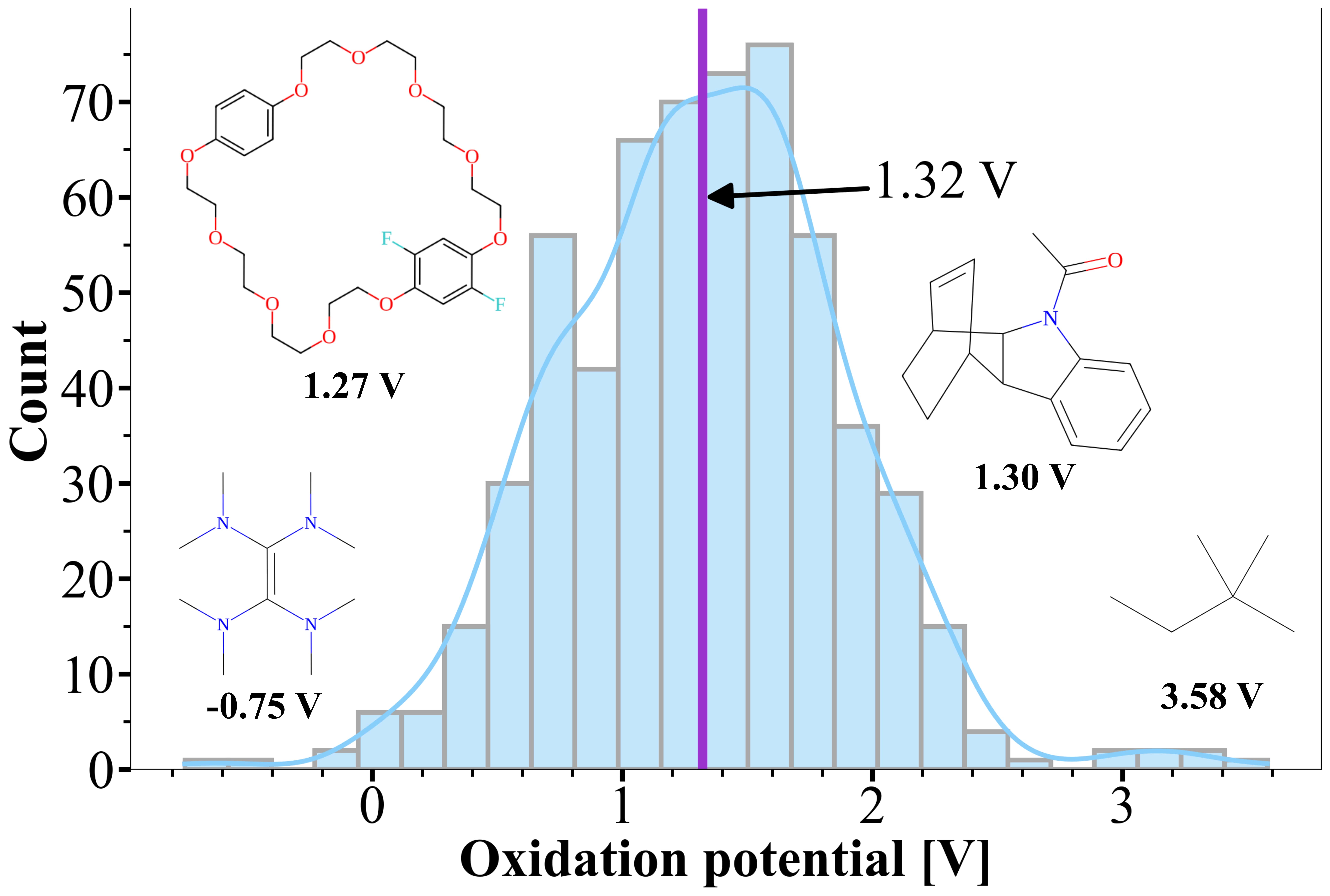}}
\caption{\label{fig: Collected data set oxidation potentials distribution}Distribution of experimentally-measured oxidation potentials 
(vs. standard calomel electrode in acetonitrile) of 592 unique neutrally-charged molecules extracted from literature. Solid vertical line indicates the mean. Exemplary molecules at the extremes and near the mean of the distribution are depicted.}
\end{figure}




\subsection{\label{sec:ML Model Performances} ML Model Performance}

\begin{figure}[h]
\resizebox{\columnwidth}{!}{\includegraphics{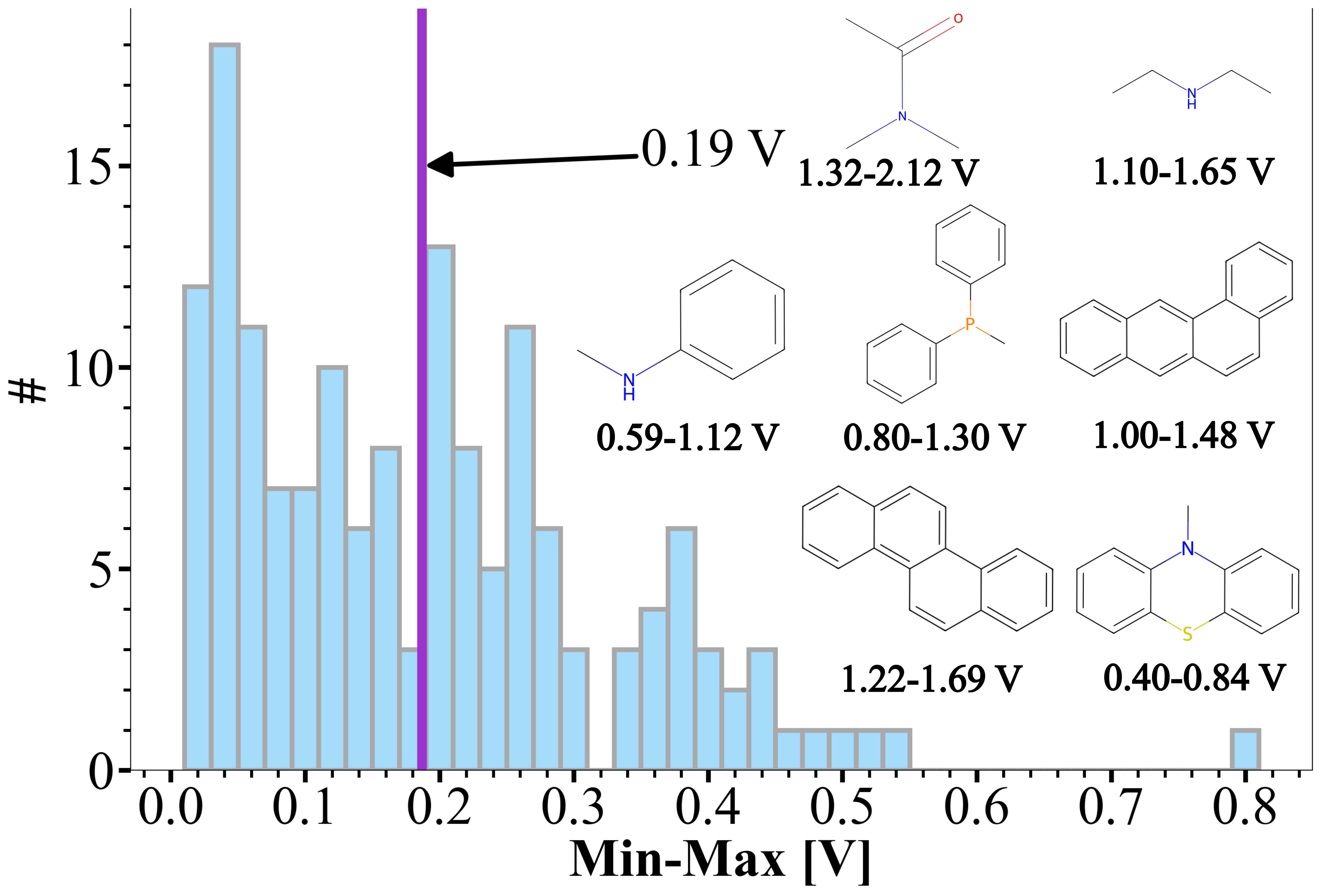}}
\caption{\label{fig: Collected data set oxidation potentials min max ranges}Distribution of min-max ranges of oxidation potentials. 
(vs. standard calomel electrode in acetonitrile) for 155 molecules with multiple entries.
Solid vertical line indicates the mean.
Seven molecules with the greatest deviations are shown.}
\end{figure}

The performances of the XGBR and KRR models were assessed by their MAE and their coefficients of determination, R$^2$. 
A target accuracy for the MAE was established as 0.2 V, which was deemed to appropriately represent experimental uncertainties since the average of the min-max range of oxidation potentials of molecules measured in multiple studies is 0.19 V (\autoref{fig: Collected data set oxidation potentials min max ranges}). 
By assessing these models by these metrics on the test set, the best performance on the out-of-sample test set was observed for the XGBR model trained on ACSF, XTB, MORDRED (MAE$_\textrm{{test}}=$ 0.15 V, R$^2$$_\textrm{{test}}=$ 0.80), followed by ACSF, XTB; ACSF; ACSF, MORDRED (see \textit{Section C} of Supplementary Information for actual vs predicted oxidation potentials of the test set). 
Similarly, the KRR model trained on the SLATM representation yields the lowest test set error (MAE$_\textrm{{test}}=$ 0.15 V; R$^2$$_\textrm{{test}}=$ 0.83) (\autoref{fig: SLATM KRR test set performance}), followed by SOAP, then ACSF (see \textit{Section C} of Supplementary Information for actual vs predicted oxidation potentials of the test set). 
From these XGBR and KRR models, the KRR model trained on SLATM achieves the best performance on the test set as it achieved the greatest R$^2$ value and the lowest MAE.\\

\begin{figure*}
     \centering
     \subfloat[\label{fig: SLATM KRR test set performance}]{\includegraphics[width=0.48\textwidth]{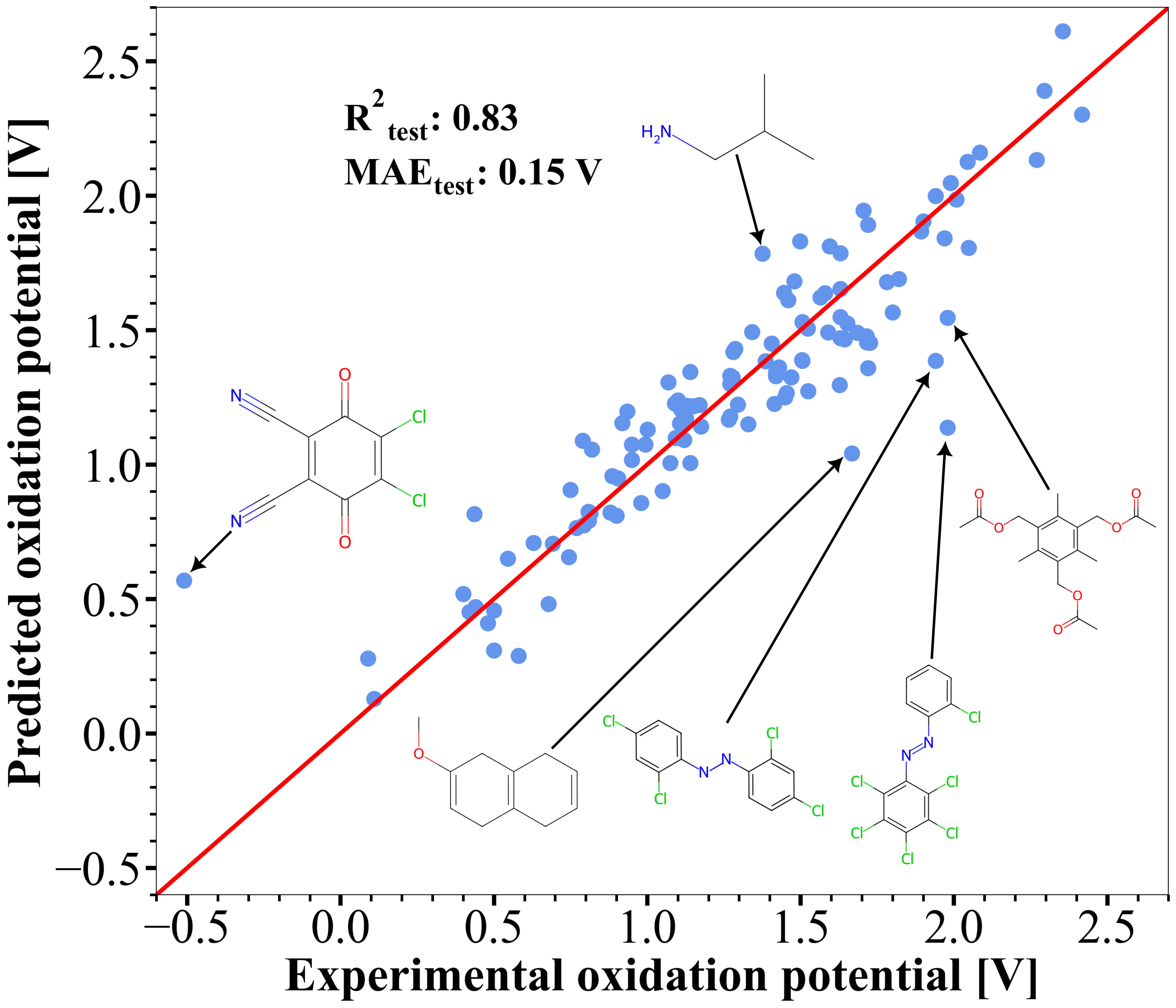}}
     \hfill
     \centering
     \subfloat[\label{fig: SLATM KRR test set performance error distribution}]{\includegraphics[width=0.48\textwidth]{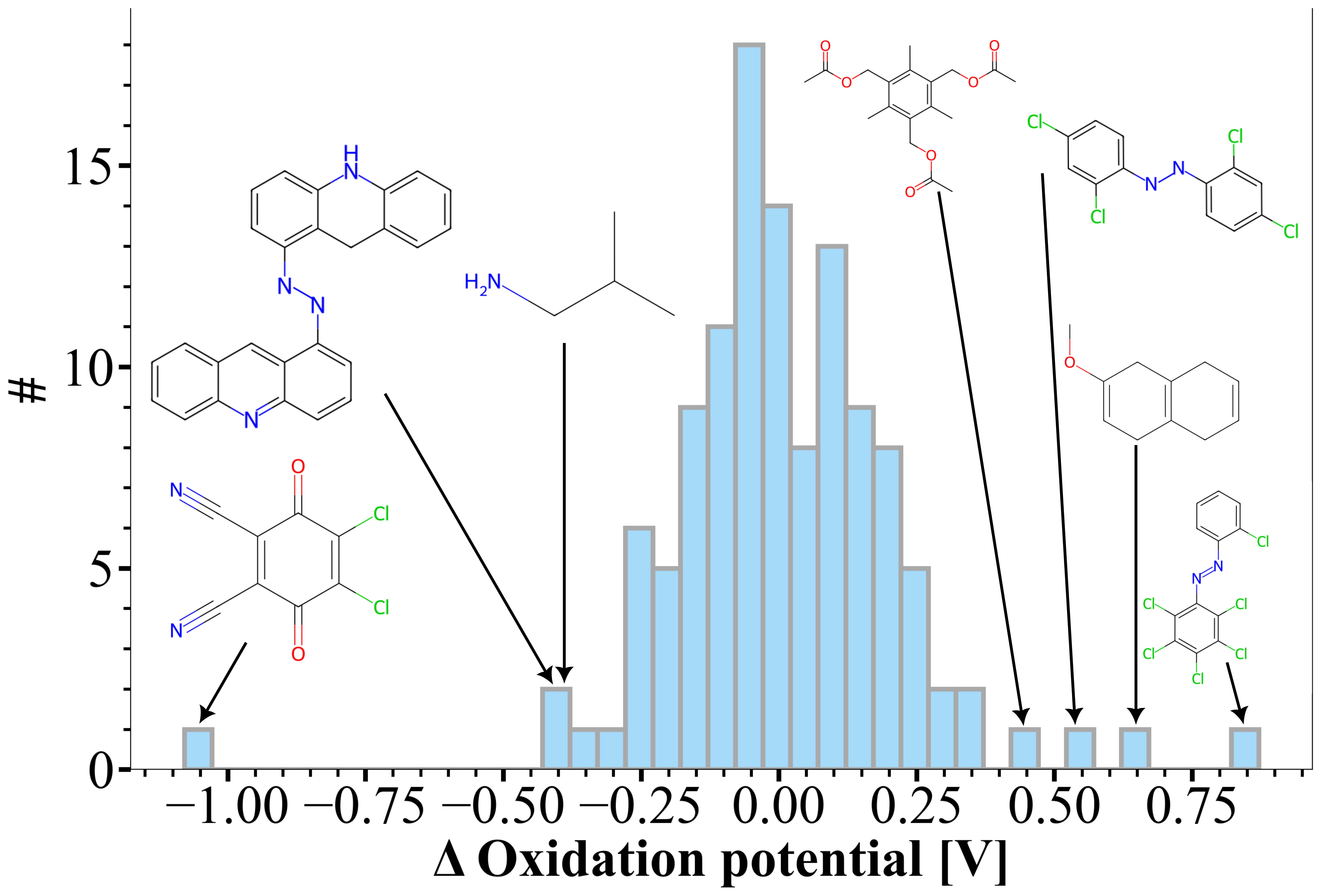}}
     \hfill
     \caption{Prediction errors of machine learning models of oxidation potentials for 119 out-of-sample molecules. Predictions were obtained by kernel ridge regression (KRR) using SLATM\cite{SLATM} as representations after training on 473 examples.  Outliers are shown as insets.
     \textbf{(a)} Scatter plot of experimental vs. predicted. 
     \textbf{(b)} Error distribution.}
\end{figure*}

\begin{figure}[h]
\resizebox{\columnwidth}{!}{\includegraphics{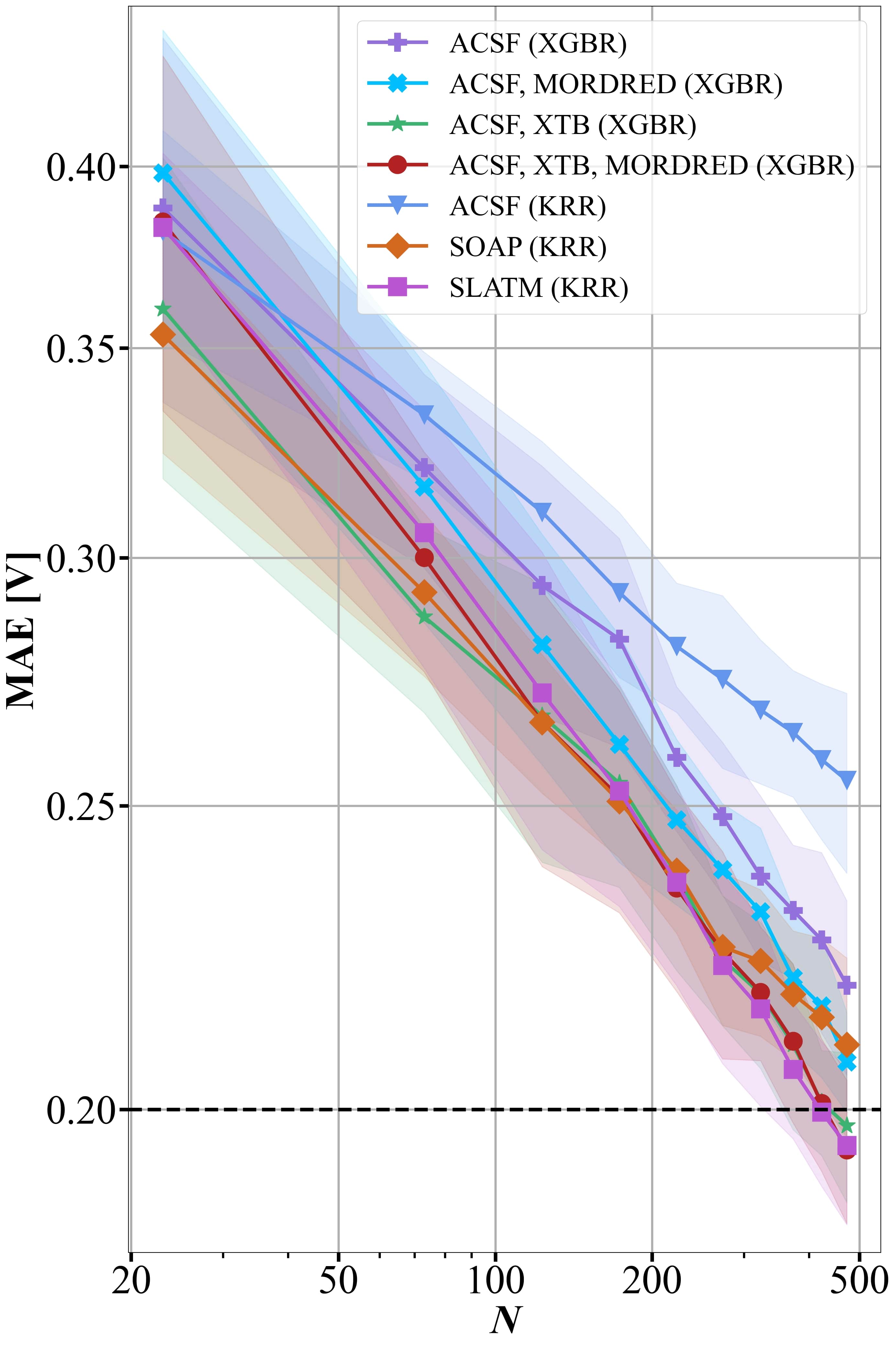}}
\caption{\label{fig: KRR and XGBoost learning curves}Learning curves: Sample test errors for predicted oxidation potentials vs training set size $N$. The shading indicates standard deviation at each number of training molecules, $N$, obtained by four-fold cross-validation for feature-based XGBR (ACSF\cite{atom_centered_symmetry_functions}; ACSF, MORDRED\cite{MORDRED}; ACSF, XTB\cite{XTB_PYTHON}; ACSF, XTB, MORDRED) and KRR (ACSF (KRR), SOAP\cite{smooth_overlap_of_atomic_positions} (KRR), SLATM\cite{SLATM} (KRR)), respectively. Dashed line indicates a target MAE of 0.2 V, corresponding to experimental uncertainty. 
}
\end{figure}

Further, the performances of the XGBR and KRR models were assessed using learning curves, which are key to evaluating the efficiency of ML models (\autoref{fig: KRR and XGBoost learning curves}). 
They show the MAEs of the various models at ten different subset sizes, $N$, of the training set, as evaluated by four-fold cross-validation and plotted on a log-log scale.
The hyperparameters of these models were optimized for the largest training set size and were fixed for the training set size.
For instance, the KRR model trained on SLATM representation reaches the target MAE of 0.2 V the fastest after training on $\sim$416 samples (70 \% of the data set), with similar performances achieved for XGBR models trained on ACSF, XTB, MORDRED and ACSF, XTB (\autoref{fig: KRR and XGBoost learning curves}).  
Compellingly, it is clear that all representations lead to systematic linear decays in the MAEs of the oxidation potentials, as is generally expected for learning curves \cite{Universal_theorem_on_learning_curves}. 
This indicates that these physics-based molecular representations and molecular descriptors are well-suited to machine learn fundamental chemical properties like oxidation potentials.
Moreover, it demonstrates that the data collected from the literature through the automated process used in this work are of sufficient quality such that experimental uncertainty in the ML-predictions can be reached with a relatively small data set. 
Further, these results suggest that the accuracy of these ML models can be systematically improved by increasing training data. 
Improvements on the automated pipeline used in this work and its application toward a larger volume of literature work may be a method to efficiently expand this data set.\\



We noticed that experimental outcomes for 155 molecules were independently reported in otherwise unrelated publications. 
The distribution of the corresponding min-max values 
is featured in \autoref{fig: Collected data set oxidation potentials min max ranges}. 
For some molecules, the deviation is considerable, and could be due to all sorts of reasons including noise from use of different experimental set-ups (e.g.~use of different reference electrodes) as well as human error.  
For example, N,N-dimethylacetamide was measured to have an oxidation potential of 1.32 V \cite{Ref_05}, or of 2.12 V\cite{Ref_24}.
To estimate which measurement values for molecules with large deviations are more likely, the fifty molecules with the largest deviations were removed from training a KRR model on SLATM (80/20 train/test random split, four-fold cross-validation for hyperparameter tuning; MAE$_\textrm{{test}}=$ 0.15 V; R$^2_\textrm{test}=$ 0.85), which was subsequently used to predict the oxidation potentials of the fifty ``suspicious" molecules (see \textit{Section C} of Supplementary Information for the performance of the KRR model on the test set).
Whichever experimental value that was closest to the predicted value was deemed to be the more likely value. 
In the case of N,N-dimethylacetamide, the ML prediction amounts to 1.90 V, statistically suggesting that 
the value of 2.12 V is closer to the truth than the value of 1.32.
This kind of scoring has been performed for all the 50 molecules left out of training (see \textit{Section D} of Supplementary Information).

\subsection{\label{sec:QM9 Analysis}Estimated Oxidation Potentials of QM9 Molecules and Descriptor-Property Analyses}

\begin{figure}[h]
\resizebox{\columnwidth}{!}{\includegraphics{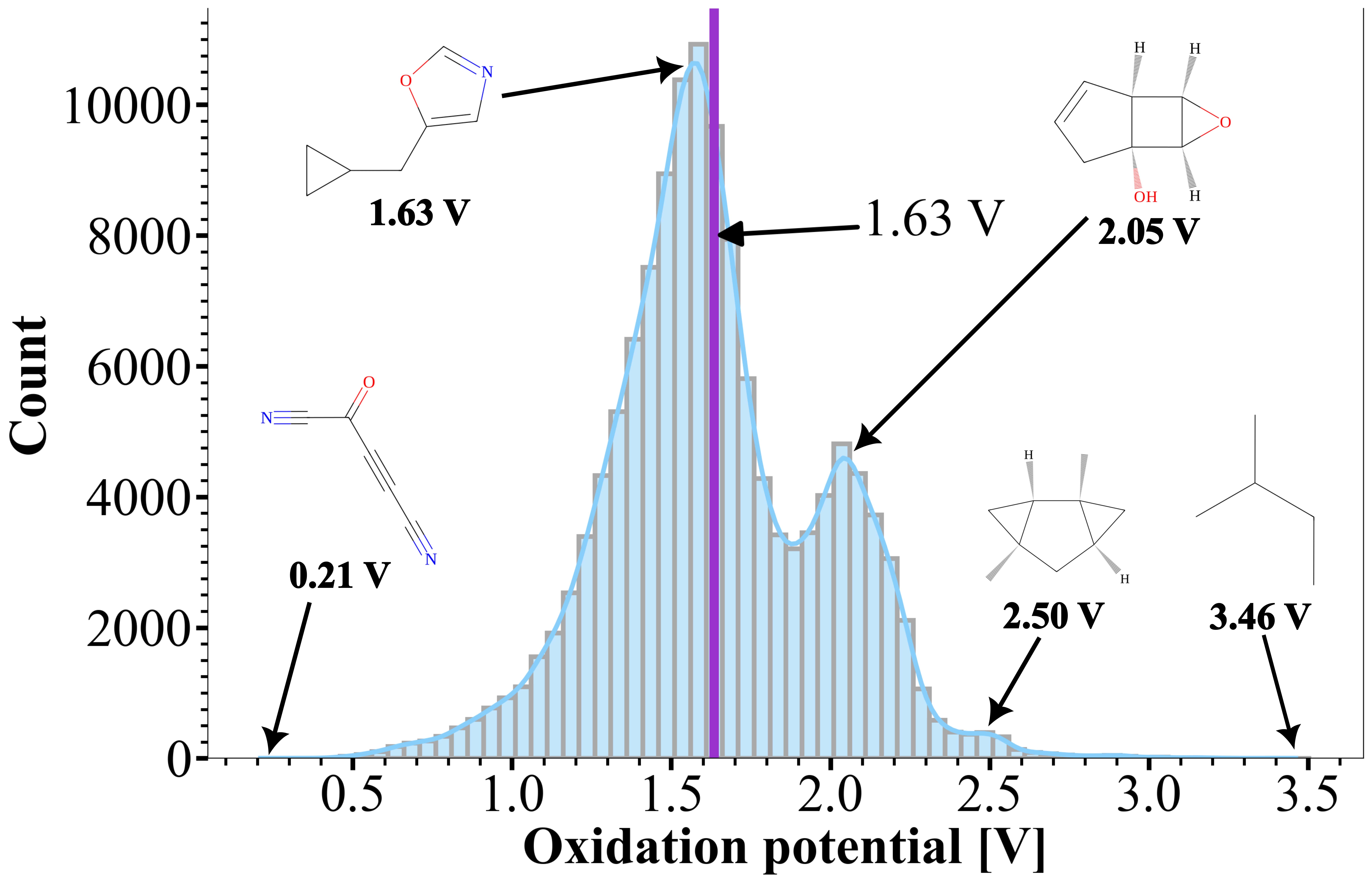}}
\caption{\label{fig: QM9 Oxidation potentials distribution}
Trimodal distribution of machine learning based predictions of oxidation potentials 
(vs standard calomel electrode in acetonitrile) 
for $\sim$132k organic molecules in the QM9 data base\cite{QM9_database, QM9_database2}. 
Model used corresponds to XGBR/ 
 ACSF and \texttt{XTB}-calculated values (green, starred in \autoref{fig: KRR and XGBoost learning curves}). Solid vertical line indicates the mean. Exemplary molecules at the extreme ends of the distribution and near the three peaks are shown as insets.}
\end{figure}

The XGBR model trained on the ACSF representations and \texttt{XTB}-calculated values (MAE$_\textrm{{test}}=$ 0.16 V, R$^2$$_\textrm{{test}}=$ 0.78) was used to estimate the oxidation potentials of 132k organic molecules contained in the QM9 database. 
QM9 does not report calculated values of oxidation potentials and as far as the author is aware, no previous work has performed a screen of the database to estimate such values. 
The geometries reported by the QM9 database were optimized in acetonitrile using \texttt{XTB} which were then used to generate the ACSF representations and inputted with the \texttt{XTB}-calculated values into the XGBR model, resulting in molecules in the QM9 database having oxidation potentials that follow a trimodal distribution, with two distinct peaks, and an average of 1.63 V (0.21--3.46 V) (\autoref{fig: QM9 Oxidation potentials distribution}).\\

\begin{figure}[h]
\resizebox{\columnwidth}{!}{\includegraphics{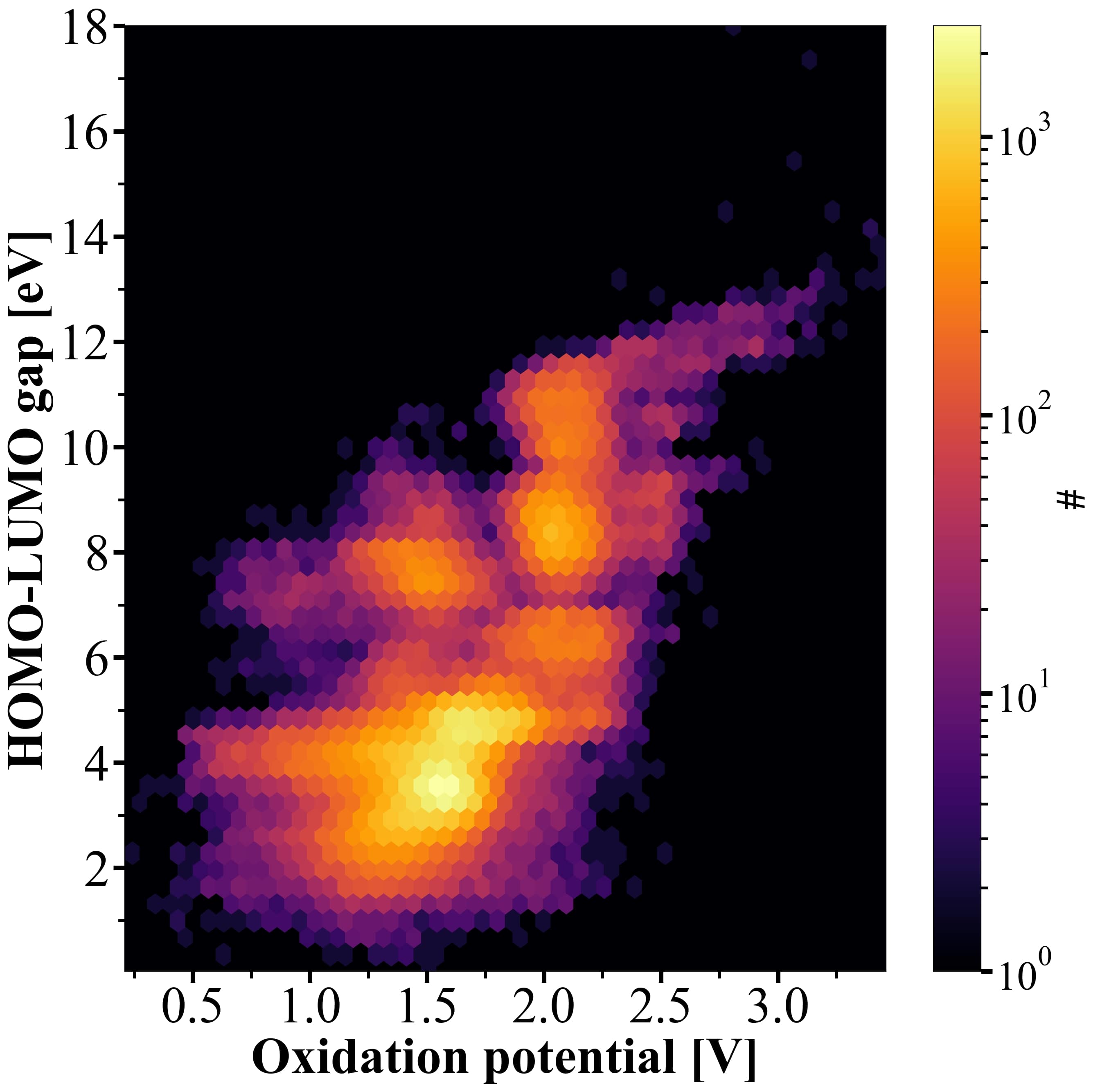}}
\caption{\label{fig: QM9 Oxidation potential vs HOMO-LUMO}Hexbin plot of machine learning model based estimated values of oxidation potentials and HOMO-LUMO gap energies 
of $\sim$132k molecules in the QM9 data set\cite{QM9_database, QM9_database2}. Color bar indicates the density of samples in each bin.}
\end{figure}

The oxidation potentials of the molecules are correlated with their corresponding \texttt{XTB}-estimated values of their HOMO-LUMO energy gaps (\autoref{fig: QM9 Oxidation potential vs HOMO-LUMO}) and solvation free energies calculated for single conformers (see \textit{Section G} of Supplementary Information for hexbin plot of oxidation potentials and solvation free energies of QM9 molecules) in acetonitrile because these are two fundamental properties of a molecule that determine its propensity to accept or donate an electron, as well as its stability in a particular solvent. 
There appears to be no obvious correlation between oxidation potentials and the two energy values, which may suggest that more data points encompassing a greater diversity of molecules are required for a clearer trend to emerge. 
However, the samples in the scatter plots appear to be clustered in certain distributions, suggesting the presence of boundaries in chemical compound space in which small organic molecules can exist with certain combinations of oxidation potentials, HOMO-LUMO gaps, and solvation free energies.\\

\begin{figure*}
     \centering
     \subfloat[\label{fig: QM9 aliphatics and nitrogen atoms distribution}]{\includegraphics[width=0.33\textwidth]{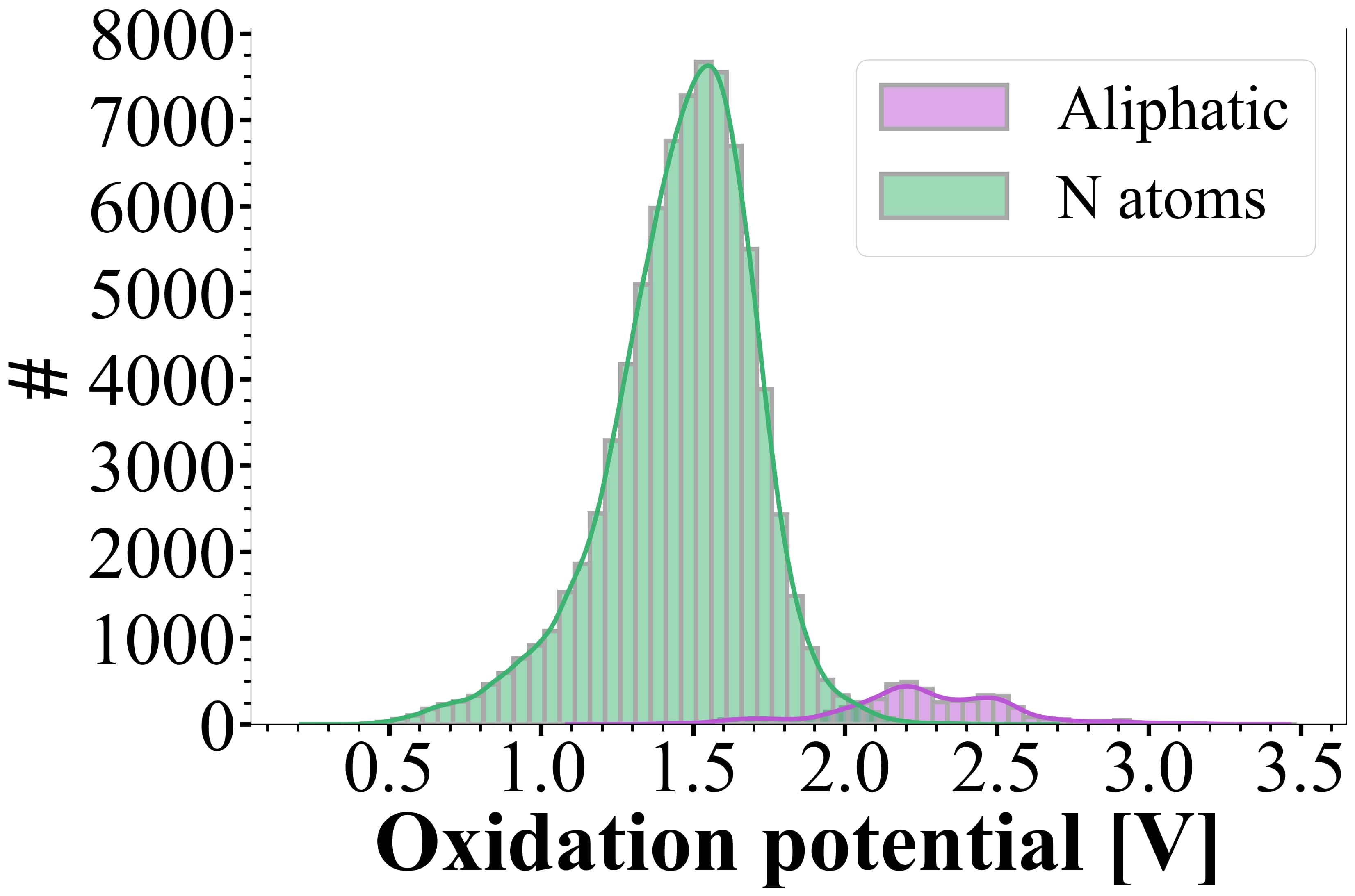}}
     \hfill
     \subfloat[\label{fig: QM9 aliphatics violin plot}]{\includegraphics[width=0.33\textwidth]{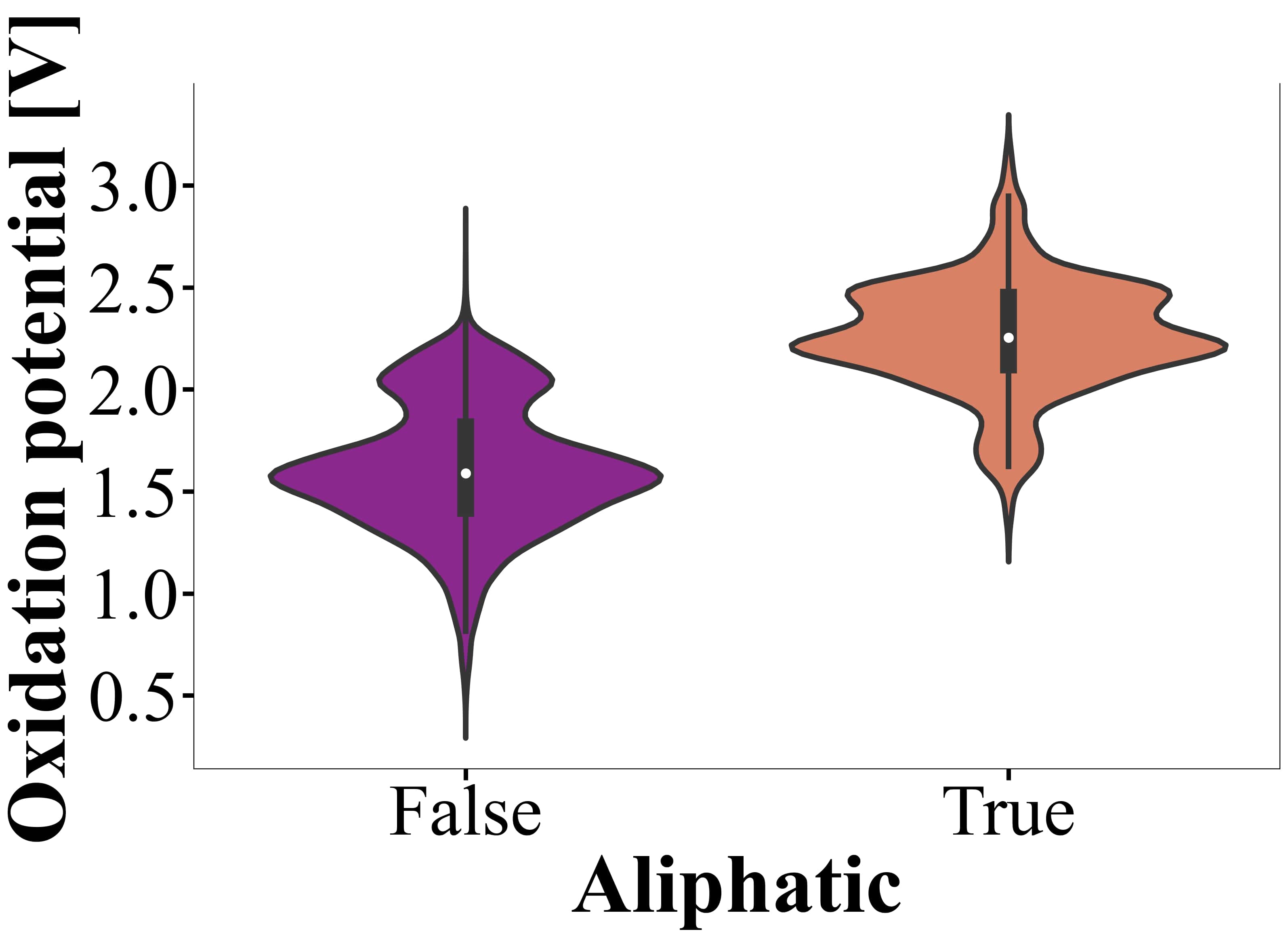}}
     \hfill
     \subfloat[\label{fig: QM9 heavy atoms violin plot}]{\includegraphics[width=0.33\textwidth]{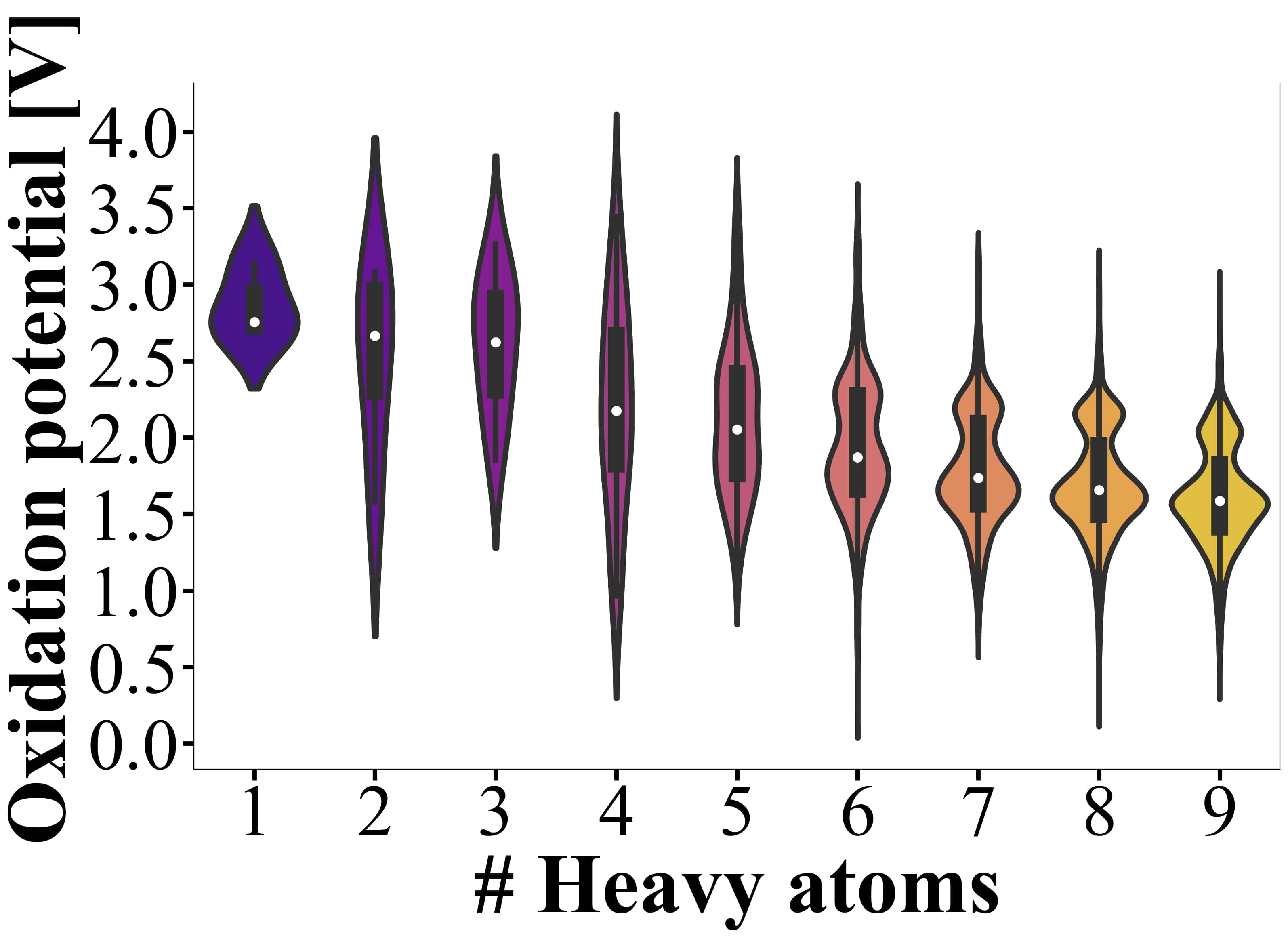}}
        \caption{Explanation of the distributions \textbf{(a)} Distributions of predicted oxidation potentials (vs. standard calomel electrode) of aliphatic and N-containing molecules in QM9 \textbf{(b)} violin plots of predicted oxidation potentials of non-aliphatic and aliphatic molecules in QM9 \textbf{(c)} violin plots of predicted oxidation potentials of molecules in QM9 classified by number of heavy atoms (excluding hydrogen). }
\end{figure*}

Previous work has shown that the distribution of HOMO-LUMO gap energies of molecules in QM9 follows a multimodal distribution with peaks that correspond to sub-distributions based on simple structural features \cite{HOMO_LUMO_gap_energies_QM9}.
To determine if the peaks in the distribution of oxidation potentials in QM9 are similarly composed of sub-distributions, a frequency analysis of functional groups and specific atoms, degree of unsaturation, and molecular types was performed using SMILES strings and substructure matching as implemented in \texttt{RDKit}\cite{RDKit} (see \textit{Section H} of Supplementary Information for full frequency analyses of functional groups and atom types).
Intriguingly, upon visual inspection of the distributions, aliphatic molecules are clustered near the peaks at $\sim$2.0, 2.5 V (\autoref{fig: QM9 aliphatics and nitrogen atoms distribution}, \ref{fig: QM9 aliphatics violin plot}).\\

However, many molecules with other structural features contribute to the peak at $\sim$1.5 V, such as molecules containing halogens, aromatics rings, amines, amides, and carbonyl groups (see \textit{Section H} of Supplementary Information for corresponding distribution plots). In particular, molecules containing nitrogens exhibit a unimodal distribution of their oxidation potentials with a peak at $\sim$1.5 V (\autoref{fig: QM9 aliphatics and nitrogen atoms distribution}).
Other trends of note include near-linear increases in oxidation potentials of molecules with greater number of rings, carbons, hydroxyl groups, ethers, and hydrogens (see \textit{Section H} of Supplementary Information for corresponding violin plots).
There also appears to be near-linear decreases in oxidation potentials of molecules with greater numbers of aldehydes, ketones, carbon-oxygen double-bonds, larger degree of unsaturation, and number of heavy atoms, with the latter displaying a particularly prominent linear relationship (\autoref{fig: QM9 heavy atoms violin plot}).

\section{\label{sec:Conclusion}Conclusion}
This work introduced an automated data-extraction pipeline involving a convolutional neural network for table detection and a large language model for the selective extraction of scientific information. 
This pipeline was utilized to extract data from 74 peer-reviewed scientific publications listing tables of experimentally-measured oxidation potentials of organic molecules, resulting in a data set of 592 unique organic molecules, their canonical SMILES, generated XYZ-coordinates, and their oxidation potentials.
ML models that reach experimental uncertainty of $\sim$0.2 V were trained on this data set, which were subsequently used to estimate the true oxidation potentials of molecules with great discrepancies across multiple measurements and determine which measurements are more reliable.
Oxidation potentials of $\sim$132k small organic molecules in the QM9 data set were also estimated using the trained ML models and correlated with simple molecular descriptors.
This analysis suggests that the oxidation potentials of these molecules depend on the number of heavy atoms and chemical compositions, in particular aliphaticity and nitrogen content.\\

These results suggest that the automated data-extraction pipelines may serve accelerated discoveries of novel molecules and materials through self-driving labs\cite{self_driving_labs}. 
More specifically, rather than generating training data from scratch, analogous pipelines can be used to train ML models for initializing the experimental planning decisions necessary to launch iterative self-driving lab campaigns.
To this end, it could be desirable to develop a deeply-connected neural network, or another algorithmic model that can achieve higher table-detection accuracies, to limit data loss. 
It might be worth investigating the incorporation of optical chemical structure recognition tools to improve a large language model's ability to recognize as molecules the bond-line structure representations and drawings. 
Further, it may be valuable to develop a large language model that is specifically trained to understand the semantics and jargon of various scientific disciplines to further improve the extraction of user-specified information.
\\

\section{\label{sec:Supplementary Information}Supplementary Information}
The supplementary information contains references of the literature sources from which data was extracted, and a table listing the samples' SMILES and experimentally-measured oxidation potentials (V, vs. SCE). 
Generated xyz-coordinates of the extracted molecules and $\sim$132k molecules in QM9, optimized in acetonitrile solvent, are provided.
Scatter plots of actual vs predicted oxidation potentials of the XGBoost and KRR models on various molecular representations are shown.
For the fifty molecules with the largest measurement deviations across multiple studies, estimations of their true oxidation potentials and the experimental values closest to these estimates are listed.
It also shows the molecules with the most positive and most negative oxidation potentials in the extracted data set, as well as for the molecules in the QM9 data set based on their ML-estimated oxidation potentials. 
Additionally, frequency analyses of the functional groups and atom types in QM9 molecules are displayed.
Sample outputs from the CNN and the LLM are displayed.
Moreover, \texttt{Python} code to construct the TableNet convolutional neural network for table detection and hyperparameters of the KRR trained on ACSF and \texttt{XTB}-calculated values are available.

\clearpage

\nocite{*}

\begin{acknowledgments} 
O.A.v.L. has received funding from the European Research Council (ERC) under the European Union’s Horizon 2020 research and innovation programme (grant agreement No.~772834).
This project has received funding from the European Union’s Horizon 2020 research and innovation programme under grant agreement No.~957189.
This research is part of the University of Toronto’s Acceleration Consortium, which receives funding from the Canada First Research Excellence Fund (CFREF).
O.A.v.L. has received support as the Ed Clark Chair of Advanced Materials and as a Canada CIFAR AI Chair.  
\end{acknowledgments}

\bibliography{References}

\begin{thebibliography}{82}%
\makeatletter
\providecommand \@ifxundefined [1]{%
 \@ifx{#1\undefined}
}%
\providecommand \@ifnum [1]{%
 \ifnum #1\expandafter \@firstoftwo
 \else \expandafter \@secondoftwo
 \fi
}%
\providecommand \@ifx [1]{%
 \ifx #1\expandafter \@firstoftwo
 \else \expandafter \@secondoftwo
 \fi
}%
\providecommand \natexlab [1]{#1}%
\providecommand \enquote  [1]{``#1''}%
\providecommand \bibnamefont  [1]{#1}%
\providecommand \bibfnamefont [1]{#1}%
\providecommand \citenamefont [1]{#1}%
\providecommand \href@noop [0]{\@secondoftwo}%
\providecommand \href [0]{\begingroup \@sanitize@url \@href}%
\providecommand \@href[1]{\@@startlink{#1}\@@href}%
\providecommand \@@href[1]{\endgroup#1\@@endlink}%
\providecommand \@sanitize@url [0]{\catcode `\\12\catcode `\$12\catcode
  `\&12\catcode `\#12\catcode `\^12\catcode `\_12\catcode `\%12\relax}%
\providecommand \@@startlink[1]{}%
\providecommand \@@endlink[0]{}%
\providecommand \url  [0]{\begingroup\@sanitize@url \@url }%
\providecommand \@url [1]{\endgroup\@href {#1}{\urlprefix }}%
\providecommand \urlprefix  [0]{URL }%
\providecommand \Eprint [0]{\href }%
\providecommand \doibase [0]{http://dx.doi.org/}%
\providecommand \selectlanguage [0]{\@gobble}%
\providecommand \bibinfo  [0]{\@secondoftwo}%
\providecommand \bibfield  [0]{\@secondoftwo}%
\providecommand \translation [1]{[#1]}%
\providecommand \BibitemOpen [0]{}%
\providecommand \bibitemStop [0]{}%
\providecommand \bibitemNoStop [0]{.\EOS\space}%
\providecommand \EOS [0]{\spacefactor3000\relax}%
\providecommand \BibitemShut  [1]{\csname bibitem#1\endcsname}%
\let\auto@bib@innerbib\@empty
\bibitem [{\citenamefont {Akhter}\ \emph {et~al.}(2019)\citenamefont {Akhter},
  \citenamefont {Pauyo},\ and\ \citenamefont
  {Khan}}]{What_is_the_difference_between_a_systematic_review_and_a_meta_analysis}%
  \BibitemOpen
  \bibfield  {author} {\bibinfo {author} {\bibfnamefont {S.}~\bibnamefont
  {Akhter}}, \bibinfo {author} {\bibfnamefont {T.}~\bibnamefont {Pauyo}}, \
  and\ \bibinfo {author} {\bibfnamefont {M.}~\bibnamefont {Khan}},\ }\href@noop
  {} {\bibfield  {journal} {\bibinfo  {journal} {Basic methods handbook for
  clinical orthopaedic research: a practical guide and case based research
  approach}\ ,\ \bibinfo {pages} {331}} (\bibinfo {year} {2019})}\BibitemShut
  {NoStop}%
\bibitem [{\citenamefont {Ahn}\ and\ \citenamefont
  {Kang}(2018)}]{introduction_to_systematic_review_and_meta_analysis}%
  \BibitemOpen
  \bibfield  {author} {\bibinfo {author} {\bibfnamefont {E.}~\bibnamefont
  {Ahn}}\ and\ \bibinfo {author} {\bibfnamefont {H.}~\bibnamefont {Kang}},\
  }\href@noop {} {\bibfield  {journal} {\bibinfo  {journal} {Korean journal of
  anesthesiology}\ }\textbf {\bibinfo {volume} {71}},\ \bibinfo {pages} {103}
  (\bibinfo {year} {2018})}\BibitemShut {NoStop}%
\bibitem [{\citenamefont
  {Owens}(2021)}]{systematic_review_brief_overview_of_methods_limitations_and_resources}%
  \BibitemOpen
  \bibfield  {author} {\bibinfo {author} {\bibfnamefont {J.~K.}\ \bibnamefont
  {Owens}},\ }\href@noop {} {\bibfield  {journal} {\bibinfo  {journal} {Nurse
  Author \& Editor}\ }\textbf {\bibinfo {volume} {31}},\ \bibinfo {pages} {69}
  (\bibinfo {year} {2021})}\BibitemShut {NoStop}%
\bibitem [{\citenamefont {B{\"u}chter}\ \emph {et~al.}(2020)\citenamefont
  {B{\"u}chter}, \citenamefont {Weise},\ and\ \citenamefont
  {Pieper}}]{development_testing_and_use_of_data_extraction_forms_in_systematic_reviews}%
  \BibitemOpen
  \bibfield  {author} {\bibinfo {author} {\bibfnamefont {R.~B.}\ \bibnamefont
  {B{\"u}chter}}, \bibinfo {author} {\bibfnamefont {A.}~\bibnamefont {Weise}},
  \ and\ \bibinfo {author} {\bibfnamefont {D.}~\bibnamefont {Pieper}},\
  }\href@noop {} {\bibfield  {journal} {\bibinfo  {journal} {BMC medical
  research methodology}\ }\textbf {\bibinfo {volume} {20}},\ \bibinfo {pages}
  {1} (\bibinfo {year} {2020})}\BibitemShut {NoStop}%
\bibitem [{\citenamefont {Bornmann}\ \emph {et~al.}(2021)\citenamefont
  {Bornmann}, \citenamefont {Haunschild},\ and\ \citenamefont
  {Mutz}}]{growth_rates_of_modern_science}%
  \BibitemOpen
  \bibfield  {author} {\bibinfo {author} {\bibfnamefont {L.}~\bibnamefont
  {Bornmann}}, \bibinfo {author} {\bibfnamefont {R.}~\bibnamefont
  {Haunschild}}, \ and\ \bibinfo {author} {\bibfnamefont {R.}~\bibnamefont
  {Mutz}},\ }\href@noop {} {\bibfield  {journal} {\bibinfo  {journal}
  {Humanities and Social Sciences Communications}\ }\textbf {\bibinfo {volume}
  {8}},\ \bibinfo {pages} {1} (\bibinfo {year} {2021})}\BibitemShut {NoStop}%
\bibitem [{\citenamefont {Larsen}\ and\ \citenamefont
  {Von~Ins}(2010)}]{the_rate_of_growth_in_scientific_publication_and_the_decline_in_coverage_provided_by_Science_Citation_Index}%
  \BibitemOpen
  \bibfield  {author} {\bibinfo {author} {\bibfnamefont {P.}~\bibnamefont
  {Larsen}}\ and\ \bibinfo {author} {\bibfnamefont {M.}~\bibnamefont
  {Von~Ins}},\ }\href@noop {} {\bibfield  {journal} {\bibinfo  {journal}
  {Scientometrics}\ }\textbf {\bibinfo {volume} {84}},\ \bibinfo {pages} {575}
  (\bibinfo {year} {2010})}\BibitemShut {NoStop}%
\bibitem [{\citenamefont {Hong}\ \emph {et~al.}(2021)\citenamefont {Hong},
  \citenamefont {Ward}, \citenamefont {Chard}, \citenamefont {Blaiszik},\ and\
  \citenamefont
  {Foster}}]{challenges_and_advances_in_information_extraction_from_scientific_literature}%
  \BibitemOpen
  \bibfield  {author} {\bibinfo {author} {\bibfnamefont {Z.}~\bibnamefont
  {Hong}}, \bibinfo {author} {\bibfnamefont {L.}~\bibnamefont {Ward}}, \bibinfo
  {author} {\bibfnamefont {K.}~\bibnamefont {Chard}}, \bibinfo {author}
  {\bibfnamefont {B.}~\bibnamefont {Blaiszik}}, \ and\ \bibinfo {author}
  {\bibfnamefont {I.}~\bibnamefont {Foster}},\ }\href@noop {} {\bibfield
  {journal} {\bibinfo  {journal} {JOM}\ }\textbf {\bibinfo {volume} {73}},\
  \bibinfo {pages} {3383} (\bibinfo {year} {2021})}\BibitemShut {NoStop}%
\bibitem [{\citenamefont {Jordan}\ and\ \citenamefont
  {Mitchell}(2015)}]{machine_learning_trends_perspectives_and_prospects}%
  \BibitemOpen
  \bibfield  {author} {\bibinfo {author} {\bibfnamefont {M.~I.}\ \bibnamefont
  {Jordan}}\ and\ \bibinfo {author} {\bibfnamefont {T.~M.}\ \bibnamefont
  {Mitchell}},\ }\href@noop {} {\bibfield  {journal} {\bibinfo  {journal}
  {Science}\ }\textbf {\bibinfo {volume} {349}},\ \bibinfo {pages} {255}
  (\bibinfo {year} {2015})}\BibitemShut {NoStop}%
\bibitem [{\citenamefont {Foody}\ \emph {et~al.}(2006)\citenamefont {Foody},
  \citenamefont {Mathur}, \citenamefont {Sanchez-Hernandez},\ and\
  \citenamefont
  {Boyd}}]{training_set_size_requirements_for_the_classification_of_a_specific_class}%
  \BibitemOpen
  \bibfield  {author} {\bibinfo {author} {\bibfnamefont {G.~M.}\ \bibnamefont
  {Foody}}, \bibinfo {author} {\bibfnamefont {A.}~\bibnamefont {Mathur}},
  \bibinfo {author} {\bibfnamefont {C.}~\bibnamefont {Sanchez-Hernandez}}, \
  and\ \bibinfo {author} {\bibfnamefont {D.~S.}\ \bibnamefont {Boyd}},\
  }\href@noop {} {\bibfield  {journal} {\bibinfo  {journal} {Remote Sensing of
  Environment}\ }\textbf {\bibinfo {volume} {104}},\ \bibinfo {pages} {1}
  (\bibinfo {year} {2006})}\BibitemShut {NoStop}%
\bibitem [{\citenamefont {Hashmi}\ \emph {et~al.}(2021)\citenamefont {Hashmi},
  \citenamefont {Liwicki}, \citenamefont {Stricker}, \citenamefont {Afzal},
  \citenamefont {Afzal},\ and\ \citenamefont
  {Afzal}}]{current_status_and_performance_analysis_of_table_recognition_in_document_images_with_deep_neural_networks}%
  \BibitemOpen
  \bibfield  {author} {\bibinfo {author} {\bibfnamefont {K.~A.}\ \bibnamefont
  {Hashmi}}, \bibinfo {author} {\bibfnamefont {M.}~\bibnamefont {Liwicki}},
  \bibinfo {author} {\bibfnamefont {D.}~\bibnamefont {Stricker}}, \bibinfo
  {author} {\bibfnamefont {M.~A.}\ \bibnamefont {Afzal}}, \bibinfo {author}
  {\bibfnamefont {M.~A.}\ \bibnamefont {Afzal}}, \ and\ \bibinfo {author}
  {\bibfnamefont {M.~Z.}\ \bibnamefont {Afzal}},\ }\href@noop {} {\bibfield
  {journal} {\bibinfo  {journal} {IEEE Access}\ }\textbf {\bibinfo {volume}
  {9}},\ \bibinfo {pages} {87663} (\bibinfo {year} {2021})}\BibitemShut
  {NoStop}%
\bibitem [{\citenamefont {Colter}\ \emph {et~al.}(2022)\citenamefont {Colter},
  \citenamefont {Fayazi}, \citenamefont {Benameur-El~Youbi}, \citenamefont
  {Kamp}, \citenamefont {Yu},\ and\ \citenamefont
  {Dreslinski}}]{tablext_a_combined_neural_network_and_heuristic_based_table_extractor}%
  \BibitemOpen
  \bibfield  {author} {\bibinfo {author} {\bibfnamefont {Z.}~\bibnamefont
  {Colter}}, \bibinfo {author} {\bibfnamefont {M.}~\bibnamefont {Fayazi}},
  \bibinfo {author} {\bibfnamefont {Z.}~\bibnamefont {Benameur-El~Youbi}},
  \bibinfo {author} {\bibfnamefont {S.}~\bibnamefont {Kamp}}, \bibinfo {author}
  {\bibfnamefont {S.}~\bibnamefont {Yu}}, \ and\ \bibinfo {author}
  {\bibfnamefont {R.}~\bibnamefont {Dreslinski}},\ }\href@noop {} {\bibfield
  {journal} {\bibinfo  {journal} {Array}\ }\textbf {\bibinfo {volume} {15}},\
  \bibinfo {pages} {100220} (\bibinfo {year} {2022})}\BibitemShut {NoStop}%
\bibitem [{\citenamefont {Paliwal}\ \emph {et~al.}(2019)\citenamefont
  {Paliwal}, \citenamefont {Vishwanath}, \citenamefont {Rahul}, \citenamefont
  {Sharma},\ and\ \citenamefont
  {Vig}}]{tablenet_deep_learning_model_for_end_to_end_table_detection_and_tabular_data_extraction_from_scanned_document_images}%
  \BibitemOpen
  \bibfield  {author} {\bibinfo {author} {\bibfnamefont {S.~S.}\ \bibnamefont
  {Paliwal}}, \bibinfo {author} {\bibfnamefont {D.}~\bibnamefont {Vishwanath}},
  \bibinfo {author} {\bibfnamefont {R.}~\bibnamefont {Rahul}}, \bibinfo
  {author} {\bibfnamefont {M.}~\bibnamefont {Sharma}}, \ and\ \bibinfo {author}
  {\bibfnamefont {L.}~\bibnamefont {Vig}},\ }in\ \href@noop {} {\emph {\bibinfo
  {booktitle} {2019 International Conference on Document Analysis and
  Recognition (ICDAR)}}}\ (\bibinfo {organization} {IEEE},\ \bibinfo {year}
  {2019})\ pp.\ \bibinfo {pages} {128--133}\BibitemShut {NoStop}%
\bibitem [{\citenamefont {G{\"o}bel}\ \emph {et~al.}(2013)\citenamefont
  {G{\"o}bel}, \citenamefont {Hassan}, \citenamefont {Oro},\ and\ \citenamefont
  {Orsi}}]{ICDAR_2013_table_competition}%
  \BibitemOpen
  \bibfield  {author} {\bibinfo {author} {\bibfnamefont {M.}~\bibnamefont
  {G{\"o}bel}}, \bibinfo {author} {\bibfnamefont {T.}~\bibnamefont {Hassan}},
  \bibinfo {author} {\bibfnamefont {E.}~\bibnamefont {Oro}}, \ and\ \bibinfo
  {author} {\bibfnamefont {G.}~\bibnamefont {Orsi}},\ }in\ \href@noop {} {\emph
  {\bibinfo {booktitle} {2013 12th International Conference on Document
  Analysis and Recognition}}}\ (\bibinfo {organization} {IEEE},\ \bibinfo
  {year} {2013})\ pp.\ \bibinfo {pages} {1449--1453}\BibitemShut {NoStop}%
\bibitem [{\citenamefont {Islam}\ \emph {et~al.}(2017)\citenamefont {Islam},
  \citenamefont {Islam},\ and\ \citenamefont
  {Noor}}]{a_survey_on_optical_character_recognition_system}%
  \BibitemOpen
  \bibfield  {author} {\bibinfo {author} {\bibfnamefont {N.}~\bibnamefont
  {Islam}}, \bibinfo {author} {\bibfnamefont {Z.}~\bibnamefont {Islam}}, \ and\
  \bibinfo {author} {\bibfnamefont {N.}~\bibnamefont {Noor}},\ }\href@noop {}
  {\bibfield  {journal} {\bibinfo  {journal} {arXiv preprint arXiv:1710.05703}\
  } (\bibinfo {year} {2017})}\BibitemShut {NoStop}%
\bibitem [{\citenamefont
  {Smith}(2007)}]{an_overview_of_the_tesseract_OCR_engine}%
  \BibitemOpen
  \bibfield  {author} {\bibinfo {author} {\bibfnamefont {R.}~\bibnamefont
  {Smith}},\ }in\ \href@noop {} {\emph {\bibinfo {booktitle} {Ninth
  international conference on document analysis and recognition (ICDAR
  2007)}}},\ Vol.~\bibinfo {volume} {2}\ (\bibinfo {organization} {IEEE},\
  \bibinfo {year} {2007})\ pp.\ \bibinfo {pages} {629--633}\BibitemShut
  {NoStop}%
\bibitem [{\citenamefont
  {Smith}(2013)}]{history_of_the_tesseract_ocr_engine_what_worked_and_what_did_not}%
  \BibitemOpen
  \bibfield  {author} {\bibinfo {author} {\bibfnamefont {R.~W.}\ \bibnamefont
  {Smith}},\ }in\ \href@noop {} {\emph {\bibinfo {booktitle} {Document
  Recognition and Retrieval XX}}},\ Vol.\ \bibinfo {volume} {8658}\ (\bibinfo
  {organization} {SPIE},\ \bibinfo {year} {2013})\ p.\ \bibinfo {pages}
  {865802}\BibitemShut {NoStop}%
\bibitem [{\citenamefont {Memon}\ \emph {et~al.}(2020)\citenamefont {Memon},
  \citenamefont {Sami}, \citenamefont {Khan},\ and\ \citenamefont
  {Uddin}}]{handwritten_optical_character_recognition}%
  \BibitemOpen
  \bibfield  {author} {\bibinfo {author} {\bibfnamefont {J.}~\bibnamefont
  {Memon}}, \bibinfo {author} {\bibfnamefont {M.}~\bibnamefont {Sami}},
  \bibinfo {author} {\bibfnamefont {R.~A.}\ \bibnamefont {Khan}}, \ and\
  \bibinfo {author} {\bibfnamefont {M.}~\bibnamefont {Uddin}},\ }\href@noop {}
  {\bibfield  {journal} {\bibinfo  {journal} {IEEE Access}\ }\textbf {\bibinfo
  {volume} {8}},\ \bibinfo {pages} {142642} (\bibinfo {year}
  {2020})}\BibitemShut {NoStop}%
\bibitem [{\citenamefont {Liu}\ \emph {et~al.}(2022)\citenamefont {Liu},
  \citenamefont {Chabot}, \citenamefont {Troncy}, \citenamefont {Huynh},
  \citenamefont {Labb{\'e}},\ and\ \citenamefont
  {Monnin}}]{tabular_data_semantic_understanding}%
  \BibitemOpen
  \bibfield  {author} {\bibinfo {author} {\bibfnamefont {J.}~\bibnamefont
  {Liu}}, \bibinfo {author} {\bibfnamefont {Y.}~\bibnamefont {Chabot}},
  \bibinfo {author} {\bibfnamefont {R.}~\bibnamefont {Troncy}}, \bibinfo
  {author} {\bibfnamefont {V.-P.}\ \bibnamefont {Huynh}}, \bibinfo {author}
  {\bibfnamefont {T.}~\bibnamefont {Labb{\'e}}}, \ and\ \bibinfo {author}
  {\bibfnamefont {P.}~\bibnamefont {Monnin}},\ }\href@noop {} {\bibfield
  {journal} {\bibinfo  {journal} {Journal of Web Semantics}\ ,\ \bibinfo
  {pages} {100761}} (\bibinfo {year} {2022})}\BibitemShut {NoStop}%
\bibitem [{\citenamefont {Zhao}\ \emph {et~al.}(2023)\citenamefont {Zhao},
  \citenamefont {Zhou}, \citenamefont {Li}, \citenamefont {Tang}, \citenamefont
  {Wang}, \citenamefont {Hou}, \citenamefont {Min}, \citenamefont {Zhang},
  \citenamefont {Zhang}, \citenamefont {Dong} \emph
  {et~al.}}]{a_survey_of_large_language_models}%
  \BibitemOpen
  \bibfield  {author} {\bibinfo {author} {\bibfnamefont {W.~X.}\ \bibnamefont
  {Zhao}}, \bibinfo {author} {\bibfnamefont {K.}~\bibnamefont {Zhou}}, \bibinfo
  {author} {\bibfnamefont {J.}~\bibnamefont {Li}}, \bibinfo {author}
  {\bibfnamefont {T.}~\bibnamefont {Tang}}, \bibinfo {author} {\bibfnamefont
  {X.}~\bibnamefont {Wang}}, \bibinfo {author} {\bibfnamefont {Y.}~\bibnamefont
  {Hou}}, \bibinfo {author} {\bibfnamefont {Y.}~\bibnamefont {Min}}, \bibinfo
  {author} {\bibfnamefont {B.}~\bibnamefont {Zhang}}, \bibinfo {author}
  {\bibfnamefont {J.}~\bibnamefont {Zhang}}, \bibinfo {author} {\bibfnamefont
  {Z.}~\bibnamefont {Dong}},  \emph {et~al.},\ }\href@noop {} {\bibfield
  {journal} {\bibinfo  {journal} {arXiv preprint arXiv:2303.18223}\ } (\bibinfo
  {year} {2023})}\BibitemShut {NoStop}%
\bibitem [{\citenamefont {Fan}\ \emph {et~al.}(2023)\citenamefont {Fan},
  \citenamefont {Li}, \citenamefont {Ma}, \citenamefont {Lee}, \citenamefont
  {Yu},\ and\ \citenamefont
  {Hemphill}}]{a_bibliometric_review_of_large_language_models_research_from_2017_to_2023}%
  \BibitemOpen
  \bibfield  {author} {\bibinfo {author} {\bibfnamefont {L.}~\bibnamefont
  {Fan}}, \bibinfo {author} {\bibfnamefont {L.}~\bibnamefont {Li}}, \bibinfo
  {author} {\bibfnamefont {Z.}~\bibnamefont {Ma}}, \bibinfo {author}
  {\bibfnamefont {S.}~\bibnamefont {Lee}}, \bibinfo {author} {\bibfnamefont
  {H.}~\bibnamefont {Yu}}, \ and\ \bibinfo {author} {\bibfnamefont
  {L.}~\bibnamefont {Hemphill}},\ }\href@noop {} {\bibfield  {journal}
  {\bibinfo  {journal} {arXiv preprint arXiv:2304.02020}\ } (\bibinfo {year}
  {2023})}\BibitemShut {NoStop}%
\bibitem [{\citenamefont {Flam-Shepherd}\ \emph {et~al.}(2022)\citenamefont
  {Flam-Shepherd}, \citenamefont {Zhu},\ and\ \citenamefont
  {Aspuru-Guzik}}]{language_models_can_learn_complex_molecular_distributions}%
  \BibitemOpen
  \bibfield  {author} {\bibinfo {author} {\bibfnamefont {D.}~\bibnamefont
  {Flam-Shepherd}}, \bibinfo {author} {\bibfnamefont {K.}~\bibnamefont {Zhu}},
  \ and\ \bibinfo {author} {\bibfnamefont {A.}~\bibnamefont {Aspuru-Guzik}},\
  }\href@noop {} {\bibfield  {journal} {\bibinfo  {journal} {Nature
  Communications}\ }\textbf {\bibinfo {volume} {13}},\ \bibinfo {pages} {3293}
  (\bibinfo {year} {2022})}\BibitemShut {NoStop}%
\bibitem [{\citenamefont
  {Grisoni}(2023)}]{chemical_language_models_for_de_novo_drug_design}%
  \BibitemOpen
  \bibfield  {author} {\bibinfo {author} {\bibfnamefont {F.}~\bibnamefont
  {Grisoni}},\ }\href@noop {} {\bibfield  {journal} {\bibinfo  {journal}
  {Current Opinion in Structural Biology}\ }\textbf {\bibinfo {volume} {79}},\
  \bibinfo {pages} {102527} (\bibinfo {year} {2023})}\BibitemShut {NoStop}%
\bibitem [{\citenamefont {Hocky}\ and\ \citenamefont
  {White}(2022)}]{natural_language_processing_models_that_automate_programming_will_transform_chemistry_research_and_teaching}%
  \BibitemOpen
  \bibfield  {author} {\bibinfo {author} {\bibfnamefont {G.~M.}\ \bibnamefont
  {Hocky}}\ and\ \bibinfo {author} {\bibfnamefont {A.~D.}\ \bibnamefont
  {White}},\ }\href@noop {} {\bibfield  {journal} {\bibinfo  {journal} {Digital
  discovery}\ }\textbf {\bibinfo {volume} {1}},\ \bibinfo {pages} {79}
  (\bibinfo {year} {2022})}\BibitemShut {NoStop}%
\bibitem [{\citenamefont {Jablonka}\ \emph {et~al.}(2023)\citenamefont
  {Jablonka}, \citenamefont {Schwaller}, \citenamefont {Ortega-Guerrero},\ and\
  \citenamefont
  {Smit}}]{is_GPT3_all_you_need_for_low_data_discovery_in_chemistry}%
  \BibitemOpen
  \bibfield  {author} {\bibinfo {author} {\bibfnamefont {K.~M.}\ \bibnamefont
  {Jablonka}}, \bibinfo {author} {\bibfnamefont {P.}~\bibnamefont {Schwaller}},
  \bibinfo {author} {\bibfnamefont {A.}~\bibnamefont {Ortega-Guerrero}}, \ and\
  \bibinfo {author} {\bibfnamefont {B.}~\bibnamefont {Smit}},\ }\href {\doibase
  10.26434/chemrxiv-2023-fw8n4} {\  (\bibinfo {year} {2023}),\
  10.26434/chemrxiv-2023-fw8n4}\BibitemShut {NoStop}%
\bibitem [{\citenamefont {Fu}\ \emph {et~al.}(2023)\citenamefont {Fu},
  \citenamefont {Wei}, \citenamefont {Song}, \citenamefont {Li}, \citenamefont
  {Xin}, \citenamefont {Omee}, \citenamefont {Dong}, \citenamefont
  {Siriwardane},\ and\ \citenamefont
  {Hu}}]{material_transformers_deep_learning_language_models_for_generative_materials_design}%
  \BibitemOpen
  \bibfield  {author} {\bibinfo {author} {\bibfnamefont {N.}~\bibnamefont
  {Fu}}, \bibinfo {author} {\bibfnamefont {L.}~\bibnamefont {Wei}}, \bibinfo
  {author} {\bibfnamefont {Y.}~\bibnamefont {Song}}, \bibinfo {author}
  {\bibfnamefont {Q.}~\bibnamefont {Li}}, \bibinfo {author} {\bibfnamefont
  {R.}~\bibnamefont {Xin}}, \bibinfo {author} {\bibfnamefont {S.~S.}\
  \bibnamefont {Omee}}, \bibinfo {author} {\bibfnamefont {R.}~\bibnamefont
  {Dong}}, \bibinfo {author} {\bibfnamefont {E.~M.~D.}\ \bibnamefont
  {Siriwardane}}, \ and\ \bibinfo {author} {\bibfnamefont {J.}~\bibnamefont
  {Hu}},\ }\href@noop {} {\bibfield  {journal} {\bibinfo  {journal} {Machine
  Learning: Science and Technology}\ }\textbf {\bibinfo {volume} {4}},\
  \bibinfo {pages} {015001} (\bibinfo {year} {2023})}\BibitemShut {NoStop}%
\bibitem [{\citenamefont {Swain}\ and\ \citenamefont
  {Cole}(2016)}]{chemdataextractor}%
  \BibitemOpen
  \bibfield  {author} {\bibinfo {author} {\bibfnamefont {M.~C.}\ \bibnamefont
  {Swain}}\ and\ \bibinfo {author} {\bibfnamefont {J.~M.}\ \bibnamefont
  {Cole}},\ }\href@noop {} {\bibfield  {journal} {\bibinfo  {journal} {Journal
  of chemical information and modeling}\ }\textbf {\bibinfo {volume} {56}},\
  \bibinfo {pages} {1894} (\bibinfo {year} {2016})}\BibitemShut {NoStop}%
\bibitem [{\citenamefont {OpenAI}()}]{introducing_chatgpt}%
  \BibitemOpen
  \bibfield  {author} {\bibinfo {author} {\bibnamefont {OpenAI}},\ }\href
  {https://openai.com/blog/chatgpt} {\enquote {\bibinfo {title} {Introducing
  chatgpt},}\ }\BibitemShut {NoStop}%
\bibitem [{\citenamefont {Eloundou}\ \emph {et~al.}(2023)\citenamefont
  {Eloundou}, \citenamefont {Manning}, \citenamefont {Mishkin},\ and\
  \citenamefont {Rock}}]{Gpts_are_gpts}%
  \BibitemOpen
  \bibfield  {author} {\bibinfo {author} {\bibfnamefont {T.}~\bibnamefont
  {Eloundou}}, \bibinfo {author} {\bibfnamefont {S.}~\bibnamefont {Manning}},
  \bibinfo {author} {\bibfnamefont {P.}~\bibnamefont {Mishkin}}, \ and\
  \bibinfo {author} {\bibfnamefont {D.}~\bibnamefont {Rock}},\ }\href@noop {}
  {\bibfield  {journal} {\bibinfo  {journal} {arXiv preprint arXiv:2303.10130}\
  } (\bibinfo {year} {2023})}\BibitemShut {NoStop}%
\bibitem [{\citenamefont {OpenAI}(2023)}]{openai2023gpt}%
  \BibitemOpen
  \bibfield  {author} {\bibinfo {author} {\bibfnamefont {R.}~\bibnamefont
  {OpenAI}},\ }\href@noop {} {\bibfield  {journal} {\bibinfo  {journal}
  {arXiv}\ } (\bibinfo {year} {2023})}\BibitemShut {NoStop}%
\bibitem [{\citenamefont {Koubaa}(2023)}]{GPT4_Vs_GPT3_5_a_concise_showdown}%
  \BibitemOpen
  \bibfield  {author} {\bibinfo {author} {\bibfnamefont {A.}~\bibnamefont
  {Koubaa}},\ }\href@noop {} {\  (\bibinfo {year} {2023})}\BibitemShut
  {NoStop}%
\bibitem [{\citenamefont {Shen}\ \emph {et~al.}(2023)\citenamefont {Shen},
  \citenamefont {Heacock}, \citenamefont {Elias}, \citenamefont {Hentel},
  \citenamefont {Reig}, \citenamefont {Shih},\ and\ \citenamefont
  {Moy}}]{chatgpt_and_other_large_language_models_are_double_edged_swords}%
  \BibitemOpen
  \bibfield  {author} {\bibinfo {author} {\bibfnamefont {Y.}~\bibnamefont
  {Shen}}, \bibinfo {author} {\bibfnamefont {L.}~\bibnamefont {Heacock}},
  \bibinfo {author} {\bibfnamefont {J.}~\bibnamefont {Elias}}, \bibinfo
  {author} {\bibfnamefont {K.~D.}\ \bibnamefont {Hentel}}, \bibinfo {author}
  {\bibfnamefont {B.}~\bibnamefont {Reig}}, \bibinfo {author} {\bibfnamefont
  {G.}~\bibnamefont {Shih}}, \ and\ \bibinfo {author} {\bibfnamefont
  {L.}~\bibnamefont {Moy}},\ }\href@noop {} {\enquote {\bibinfo {title}
  {Chatgpt and other large language models are double-edged swords},}\ }
  (\bibinfo {year} {2023})\BibitemShut {NoStop}%
\bibitem [{\citenamefont {Zhong}\ \emph {et~al.}(2020)\citenamefont {Zhong},
  \citenamefont {Yang}, \citenamefont {Ding},\ and\ \citenamefont
  {Jia}}]{organic_electrolytes_for_redox_flow_batteries}%
  \BibitemOpen
  \bibfield  {author} {\bibinfo {author} {\bibfnamefont {F.}~\bibnamefont
  {Zhong}}, \bibinfo {author} {\bibfnamefont {M.}~\bibnamefont {Yang}},
  \bibinfo {author} {\bibfnamefont {M.}~\bibnamefont {Ding}}, \ and\ \bibinfo
  {author} {\bibfnamefont {C.}~\bibnamefont {Jia}},\ }\href@noop {} {\bibfield
  {journal} {\bibinfo  {journal} {Frontiers in Chemistry}\ }\textbf {\bibinfo
  {volume} {8}},\ \bibinfo {pages} {451} (\bibinfo {year} {2020})}\BibitemShut
  {NoStop}%
\bibitem [{\citenamefont {de~la Cruz}\ \emph {et~al.}(2020)\citenamefont {de~la
  Cruz}, \citenamefont {Molina}, \citenamefont {Patil}, \citenamefont
  {Ventosa}, \citenamefont {Marcilla},\ and\ \citenamefont
  {Mavrandonakis}}]{new_insights_into_phenazine_based_RFBs}%
  \BibitemOpen
  \bibfield  {author} {\bibinfo {author} {\bibfnamefont {C.}~\bibnamefont
  {de~la Cruz}}, \bibinfo {author} {\bibfnamefont {A.}~\bibnamefont {Molina}},
  \bibinfo {author} {\bibfnamefont {N.}~\bibnamefont {Patil}}, \bibinfo
  {author} {\bibfnamefont {E.}~\bibnamefont {Ventosa}}, \bibinfo {author}
  {\bibfnamefont {R.}~\bibnamefont {Marcilla}}, \ and\ \bibinfo {author}
  {\bibfnamefont {A.}~\bibnamefont {Mavrandonakis}},\ }\href@noop {} {\bibfield
   {journal} {\bibinfo  {journal} {Sustainable Energy \& Fuels}\ }\textbf
  {\bibinfo {volume} {4}},\ \bibinfo {pages} {5513} (\bibinfo {year}
  {2020})}\BibitemShut {NoStop}%
\bibitem [{\citenamefont {Cao}\ \emph {et~al.}(2020)\citenamefont {Cao},
  \citenamefont {Tian}, \citenamefont {Xu},\ and\ \citenamefont
  {Wang}}]{organic_flow_batteries_recent_progress_and_perspectives}%
  \BibitemOpen
  \bibfield  {author} {\bibinfo {author} {\bibfnamefont {J.}~\bibnamefont
  {Cao}}, \bibinfo {author} {\bibfnamefont {J.}~\bibnamefont {Tian}}, \bibinfo
  {author} {\bibfnamefont {J.}~\bibnamefont {Xu}}, \ and\ \bibinfo {author}
  {\bibfnamefont {Y.}~\bibnamefont {Wang}},\ }\href@noop {} {\bibfield
  {journal} {\bibinfo  {journal} {Energy \& Fuels}\ }\textbf {\bibinfo {volume}
  {34}},\ \bibinfo {pages} {13384} (\bibinfo {year} {2020})}\BibitemShut
  {NoStop}%
\bibitem [{\citenamefont {Li}\ \emph {et~al.}(2020)\citenamefont {Li},
  \citenamefont {Rhodes}, \citenamefont {Cabrera-Pardo},\ and\ \citenamefont
  {Minteer}}]{recent_advancements_in_rational_design_of_nonaqueous_RFBs}%
  \BibitemOpen
  \bibfield  {author} {\bibinfo {author} {\bibfnamefont {M.}~\bibnamefont
  {Li}}, \bibinfo {author} {\bibfnamefont {Z.}~\bibnamefont {Rhodes}}, \bibinfo
  {author} {\bibfnamefont {J.~R.}\ \bibnamefont {Cabrera-Pardo}}, \ and\
  \bibinfo {author} {\bibfnamefont {S.~D.}\ \bibnamefont {Minteer}},\
  }\href@noop {} {\bibfield  {journal} {\bibinfo  {journal} {Sustainable Energy
  \& Fuels}\ }\textbf {\bibinfo {volume} {4}},\ \bibinfo {pages} {4370}
  (\bibinfo {year} {2020})}\BibitemShut {NoStop}%
\bibitem [{\citenamefont {Ramakrishnan}\ \emph {et~al.}(2014)\citenamefont
  {Ramakrishnan}, \citenamefont {Dral}, \citenamefont {Rupp},\ and\
  \citenamefont {Von~Lilienfeld}}]{QM9_database}%
  \BibitemOpen
  \bibfield  {author} {\bibinfo {author} {\bibfnamefont {R.}~\bibnamefont
  {Ramakrishnan}}, \bibinfo {author} {\bibfnamefont {P.~O.}\ \bibnamefont
  {Dral}}, \bibinfo {author} {\bibfnamefont {M.}~\bibnamefont {Rupp}}, \ and\
  \bibinfo {author} {\bibfnamefont {O.~A.}\ \bibnamefont {Von~Lilienfeld}},\
  }\href@noop {} {\bibfield  {journal} {\bibinfo  {journal} {Scientific data}\
  }\textbf {\bibinfo {volume} {1}},\ \bibinfo {pages} {1} (\bibinfo {year}
  {2014})}\BibitemShut {NoStop}%
\bibitem [{\citenamefont {Marenich}\ \emph {et~al.}(2014)\citenamefont
  {Marenich}, \citenamefont {Ho}, \citenamefont {Coote}, \citenamefont
  {Cramer},\ and\ \citenamefont {Truhlar}}]{computational_electrochemistry}%
  \BibitemOpen
  \bibfield  {author} {\bibinfo {author} {\bibfnamefont {A.~V.}\ \bibnamefont
  {Marenich}}, \bibinfo {author} {\bibfnamefont {J.}~\bibnamefont {Ho}},
  \bibinfo {author} {\bibfnamefont {M.~L.}\ \bibnamefont {Coote}}, \bibinfo
  {author} {\bibfnamefont {C.~J.}\ \bibnamefont {Cramer}}, \ and\ \bibinfo
  {author} {\bibfnamefont {D.~G.}\ \bibnamefont {Truhlar}},\ }\href@noop {}
  {\bibfield  {journal} {\bibinfo  {journal} {Physical Chemistry Chemical
  Physics}\ }\textbf {\bibinfo {volume} {16}},\ \bibinfo {pages} {15068}
  (\bibinfo {year} {2014})}\BibitemShut {NoStop}%
\bibitem [{\citenamefont {Baik}\ and\ \citenamefont
  {Friesner}(2002)}]{DFT_computation_of_redox_potentials_in_solution}%
  \BibitemOpen
  \bibfield  {author} {\bibinfo {author} {\bibfnamefont {M.-H.}\ \bibnamefont
  {Baik}}\ and\ \bibinfo {author} {\bibfnamefont {R.~A.}\ \bibnamefont
  {Friesner}},\ }\href@noop {} {\bibfield  {journal} {\bibinfo  {journal} {The
  Journal of Physical Chemistry A}\ }\textbf {\bibinfo {volume} {106}},\
  \bibinfo {pages} {7407} (\bibinfo {year} {2002})}\BibitemShut {NoStop}%
\bibitem [{\citenamefont {Bachman}\ \emph {et~al.}(2014)\citenamefont
  {Bachman}, \citenamefont {Curtiss},\ and\ \citenamefont
  {Assary}}]{Anthraquinone_redox_chemistry_DFT}%
  \BibitemOpen
  \bibfield  {author} {\bibinfo {author} {\bibfnamefont {J.~E.}\ \bibnamefont
  {Bachman}}, \bibinfo {author} {\bibfnamefont {L.~A.}\ \bibnamefont
  {Curtiss}}, \ and\ \bibinfo {author} {\bibfnamefont {R.~S.}\ \bibnamefont
  {Assary}},\ }\href@noop {} {\bibfield  {journal} {\bibinfo  {journal} {The
  Journal of Physical Chemistry A}\ }\textbf {\bibinfo {volume} {118}},\
  \bibinfo {pages} {8852} (\bibinfo {year} {2014})}\BibitemShut {NoStop}%
\bibitem [{\citenamefont {Hruska}\ \emph {et~al.}(2022)\citenamefont {Hruska},
  \citenamefont {Gale},\ and\ \citenamefont
  {Liu}}]{ML_corrections_to_DFT_calculated_redox_potentials}%
  \BibitemOpen
  \bibfield  {author} {\bibinfo {author} {\bibfnamefont {E.}~\bibnamefont
  {Hruska}}, \bibinfo {author} {\bibfnamefont {A.}~\bibnamefont {Gale}}, \ and\
  \bibinfo {author} {\bibfnamefont {F.}~\bibnamefont {Liu}},\ }\href@noop {}
  {\bibfield  {journal} {\bibinfo  {journal} {Journal of Chemical Theory and
  Computation}\ }\textbf {\bibinfo {volume} {18}},\ \bibinfo {pages} {1096}
  (\bibinfo {year} {2022})}\BibitemShut {NoStop}%
\bibitem [{\citenamefont {Zhang}\ \emph {et~al.}(2017)\citenamefont {Zhang},
  \citenamefont {Zhang}, \citenamefont {Wu}, \citenamefont {Wang},\ and\
  \citenamefont {van~der Spoel}}]{implicit_vs_explicit_solvent_models}%
  \BibitemOpen
  \bibfield  {author} {\bibinfo {author} {\bibfnamefont {J.}~\bibnamefont
  {Zhang}}, \bibinfo {author} {\bibfnamefont {H.}~\bibnamefont {Zhang}},
  \bibinfo {author} {\bibfnamefont {T.}~\bibnamefont {Wu}}, \bibinfo {author}
  {\bibfnamefont {Q.}~\bibnamefont {Wang}}, \ and\ \bibinfo {author}
  {\bibfnamefont {D.}~\bibnamefont {van~der Spoel}},\ }\href@noop {} {\bibfield
   {journal} {\bibinfo  {journal} {Journal of chemical theory and computation}\
  }\textbf {\bibinfo {volume} {13}},\ \bibinfo {pages} {1034} (\bibinfo {year}
  {2017})}\BibitemShut {NoStop}%
\bibitem [{\citenamefont {Cramer}\ \emph {et~al.}(1999)\citenamefont {Cramer},
  \citenamefont {Truhlar} \emph {et~al.}}]{implicit_solvation_model_review}%
  \BibitemOpen
  \bibfield  {author} {\bibinfo {author} {\bibfnamefont {C.~J.}\ \bibnamefont
  {Cramer}}, \bibinfo {author} {\bibfnamefont {D.~G.}\ \bibnamefont {Truhlar}},
   \emph {et~al.},\ }\href@noop {} {\bibfield  {journal} {\bibinfo  {journal}
  {Chemical Reviews}\ }\textbf {\bibinfo {volume} {99}},\ \bibinfo {pages}
  {2161} (\bibinfo {year} {1999})}\BibitemShut {NoStop}%
\bibitem [{\citenamefont {Zhang}\ and\ \citenamefont
  {Xu}(2020)}]{machine_learning_properties_of_electrolyte_additives}%
  \BibitemOpen
  \bibfield  {author} {\bibinfo {author} {\bibfnamefont {Y.}~\bibnamefont
  {Zhang}}\ and\ \bibinfo {author} {\bibfnamefont {X.}~\bibnamefont {Xu}},\
  }\href@noop {} {\bibfield  {journal} {\bibinfo  {journal} {Industrial \&
  Engineering Chemistry Research}\ }\textbf {\bibinfo {volume} {60}},\ \bibinfo
  {pages} {343} (\bibinfo {year} {2020})}\BibitemShut {NoStop}%
\bibitem [{\citenamefont {Ghule}\ \emph {et~al.}(2022)\citenamefont {Ghule},
  \citenamefont {Dash}, \citenamefont {Bagchi}, \citenamefont {Joshi},\ and\
  \citenamefont
  {Vanka}}]{predicting_redox_potentials_of_phenazine_derivatives}%
  \BibitemOpen
  \bibfield  {author} {\bibinfo {author} {\bibfnamefont {S.}~\bibnamefont
  {Ghule}}, \bibinfo {author} {\bibfnamefont {S.~R.}\ \bibnamefont {Dash}},
  \bibinfo {author} {\bibfnamefont {S.}~\bibnamefont {Bagchi}}, \bibinfo
  {author} {\bibfnamefont {K.}~\bibnamefont {Joshi}}, \ and\ \bibinfo {author}
  {\bibfnamefont {K.}~\bibnamefont {Vanka}},\ }\href@noop {} {\bibfield
  {journal} {\bibinfo  {journal} {ACS omega}\ }\textbf {\bibinfo {volume}
  {7}},\ \bibinfo {pages} {11742} (\bibinfo {year} {2022})}\BibitemShut
  {NoStop}%
\bibitem [{\citenamefont {Wang}\ and\ \citenamefont
  {Cheng}(2022)}]{automated_workflow_for_computation_of_redox_potentials}%
  \BibitemOpen
  \bibfield  {author} {\bibinfo {author} {\bibfnamefont {F.}~\bibnamefont
  {Wang}}\ and\ \bibinfo {author} {\bibfnamefont {J.}~\bibnamefont {Cheng}},\
  }\href@noop {} {\bibfield  {journal} {\bibinfo  {journal} {The Journal of
  Chemical Physics}\ }\textbf {\bibinfo {volume} {157}},\ \bibinfo {pages}
  {024103} (\bibinfo {year} {2022})}\BibitemShut {NoStop}%
\bibitem [{\citenamefont {Allam}\ \emph {et~al.}(2020)\citenamefont {Allam},
  \citenamefont {Kuramshin}, \citenamefont {Stoichev}, \citenamefont {Cho},
  \citenamefont {Lee},\ and\ \citenamefont
  {Jang}}]{molecular_structure_redox_potential_relationship_for_organic_electrode_materials}%
  \BibitemOpen
  \bibfield  {author} {\bibinfo {author} {\bibfnamefont {O.}~\bibnamefont
  {Allam}}, \bibinfo {author} {\bibfnamefont {R.}~\bibnamefont {Kuramshin}},
  \bibinfo {author} {\bibfnamefont {Z.}~\bibnamefont {Stoichev}}, \bibinfo
  {author} {\bibfnamefont {B.}~\bibnamefont {Cho}}, \bibinfo {author}
  {\bibfnamefont {S.}~\bibnamefont {Lee}}, \ and\ \bibinfo {author}
  {\bibfnamefont {S.}~\bibnamefont {Jang}},\ }\href@noop {} {\bibfield
  {journal} {\bibinfo  {journal} {Materials Today Energy}\ }\textbf {\bibinfo
  {volume} {17}},\ \bibinfo {pages} {100482} (\bibinfo {year}
  {2020})}\BibitemShut {NoStop}%
\bibitem [{Pyt()}]{Pytesseract}%
  \BibitemOpen
  \href@noop {} {\enquote {\bibinfo {title} {pytesseract --- pypi.org},}\
  }\bibinfo {howpublished}
  {\url{https://pypi.org/project/pytesseract/}}\BibitemShut {NoStop}%
\bibitem [{\citenamefont {Huang}\ \emph {et~al.}(2017)\citenamefont {Huang},
  \citenamefont {Liu}, \citenamefont {Van Der~Maaten},\ and\ \citenamefont
  {Weinberger}}]{densely_connected_convolutional_networks}%
  \BibitemOpen
  \bibfield  {author} {\bibinfo {author} {\bibfnamefont {G.}~\bibnamefont
  {Huang}}, \bibinfo {author} {\bibfnamefont {Z.}~\bibnamefont {Liu}}, \bibinfo
  {author} {\bibfnamefont {L.}~\bibnamefont {Van Der~Maaten}}, \ and\ \bibinfo
  {author} {\bibfnamefont {K.~Q.}\ \bibnamefont {Weinberger}},\ }in\ \href@noop
  {} {\emph {\bibinfo {booktitle} {Proceedings of the IEEE conference on
  computer vision and pattern recognition}}}\ (\bibinfo {year} {2017})\ pp.\
  \bibinfo {pages} {4700--4708}\BibitemShut {NoStop}%
\bibitem [{\citenamefont {Fang}\ \emph {et~al.}(2012)\citenamefont {Fang},
  \citenamefont {Tao}, \citenamefont {Tang}, \citenamefont {Qiu},\ and\
  \citenamefont
  {Liu}}]{dataset_ground_truth_and_performance_metrics_for_table_detection_evaluaiton}%
  \BibitemOpen
  \bibfield  {author} {\bibinfo {author} {\bibfnamefont {J.}~\bibnamefont
  {Fang}}, \bibinfo {author} {\bibfnamefont {X.}~\bibnamefont {Tao}}, \bibinfo
  {author} {\bibfnamefont {Z.}~\bibnamefont {Tang}}, \bibinfo {author}
  {\bibfnamefont {R.}~\bibnamefont {Qiu}}, \ and\ \bibinfo {author}
  {\bibfnamefont {Y.}~\bibnamefont {Liu}},\ }in\ \href@noop {} {\emph {\bibinfo
  {booktitle} {2012 10th IAPR International Workshop on Document Analysis
  Systems}}}\ (\bibinfo {organization} {IEEE},\ \bibinfo {year} {2012})\ pp.\
  \bibinfo {pages} {445--449}\BibitemShut {NoStop}%
\bibitem [{pdf()}]{pdf2image}%
  \BibitemOpen
  \href@noop {} {\enquote {\bibinfo {title} {pdf2image --- pypi.org},}\
  }\bibinfo {howpublished}
  {\url{https://pypi.org/project/pdf2image/}}\BibitemShut {NoStop}%
\bibitem [{\citenamefont {Elgrishi}\ \emph {et~al.}(2018)\citenamefont
  {Elgrishi}, \citenamefont {Rountree}, \citenamefont {McCarthy}, \citenamefont
  {Rountree}, \citenamefont {Eisenhart},\ and\ \citenamefont
  {Dempsey}}]{beginners_guide_to_CV}%
  \BibitemOpen
  \bibfield  {author} {\bibinfo {author} {\bibfnamefont {N.}~\bibnamefont
  {Elgrishi}}, \bibinfo {author} {\bibfnamefont {K.~J.}\ \bibnamefont
  {Rountree}}, \bibinfo {author} {\bibfnamefont {B.~D.}\ \bibnamefont
  {McCarthy}}, \bibinfo {author} {\bibfnamefont {E.~S.}\ \bibnamefont
  {Rountree}}, \bibinfo {author} {\bibfnamefont {T.~T.}\ \bibnamefont
  {Eisenhart}}, \ and\ \bibinfo {author} {\bibfnamefont {J.~L.}\ \bibnamefont
  {Dempsey}},\ }\href@noop {} {\bibfield  {journal} {\bibinfo  {journal}
  {Journal of chemical education}\ }\textbf {\bibinfo {volume} {95}},\ \bibinfo
  {pages} {197} (\bibinfo {year} {2018})}\BibitemShut {NoStop}%
\bibitem [{\citenamefont {Lemm}\ \emph {et~al.}(2021)\citenamefont {Lemm},
  \citenamefont {von Rudorff},\ and\ \citenamefont {von Lilienfeld}}]{LERULI}%
  \BibitemOpen
  \bibfield  {author} {\bibinfo {author} {\bibfnamefont {D.}~\bibnamefont
  {Lemm}}, \bibinfo {author} {\bibfnamefont {G.}~\bibnamefont {von Rudorff}}, \
  and\ \bibinfo {author} {\bibfnamefont {A.}~\bibnamefont {von Lilienfeld}},\
  }\href@noop {} {\enquote {\bibinfo {title} {Leruli. com, online molecular
  property predictions in real time and for free},}\ } (\bibinfo {year}
  {2021})\BibitemShut {NoStop}%
\bibitem [{\citenamefont {Weininger}(1988)}]{SMILES}%
  \BibitemOpen
  \bibfield  {author} {\bibinfo {author} {\bibfnamefont {D.}~\bibnamefont
  {Weininger}},\ }\href@noop {} {\bibfield  {journal} {\bibinfo  {journal}
  {Journal of chemical information and computer sciences}\ }\textbf {\bibinfo
  {volume} {28}},\ \bibinfo {pages} {31} (\bibinfo {year} {1988})}\BibitemShut
  {NoStop}%
\bibitem [{\citenamefont {Landrum}\ \emph {et~al.}(2013)\citenamefont {Landrum}
  \emph {et~al.}}]{RDKit}%
  \BibitemOpen
  \bibfield  {author} {\bibinfo {author} {\bibfnamefont {G.}~\bibnamefont
  {Landrum}} \emph {et~al.},\ }\href@noop {} {\bibfield  {journal} {\bibinfo
  {journal} {Greg Landrum}\ }\textbf {\bibinfo {volume} {8}},\ \bibinfo {pages}
  {31} (\bibinfo {year} {2013})}\BibitemShut {NoStop}%
\bibitem [{\citenamefont {Bannwarth}\ \emph {et~al.}(2021)\citenamefont
  {Bannwarth}, \citenamefont {Caldeweyher}, \citenamefont {Ehlert},
  \citenamefont {Hansen}, \citenamefont {Pracht}, \citenamefont {Seibert},
  \citenamefont {Spicher},\ and\ \citenamefont {Grimme}}]{XTB_METHODS}%
  \BibitemOpen
  \bibfield  {author} {\bibinfo {author} {\bibfnamefont {C.}~\bibnamefont
  {Bannwarth}}, \bibinfo {author} {\bibfnamefont {E.}~\bibnamefont
  {Caldeweyher}}, \bibinfo {author} {\bibfnamefont {S.}~\bibnamefont {Ehlert}},
  \bibinfo {author} {\bibfnamefont {A.}~\bibnamefont {Hansen}}, \bibinfo
  {author} {\bibfnamefont {P.}~\bibnamefont {Pracht}}, \bibinfo {author}
  {\bibfnamefont {J.}~\bibnamefont {Seibert}}, \bibinfo {author} {\bibfnamefont
  {S.}~\bibnamefont {Spicher}}, \ and\ \bibinfo {author} {\bibfnamefont
  {S.}~\bibnamefont {Grimme}},\ }\href@noop {} {\bibfield  {journal} {\bibinfo
  {journal} {Wiley Interdisciplinary Reviews: Computational Molecular Science}\
  }\textbf {\bibinfo {volume} {11}},\ \bibinfo {pages} {e1493} (\bibinfo {year}
  {2021})}\BibitemShut {NoStop}%
\bibitem [{\citenamefont {Grimme}\ \emph {et~al.}(2017)\citenamefont {Grimme},
  \citenamefont {Bannwarth},\ and\ \citenamefont {Shushkov}}]{XTB_PYTHON}%
  \BibitemOpen
  \bibfield  {author} {\bibinfo {author} {\bibfnamefont {S.}~\bibnamefont
  {Grimme}}, \bibinfo {author} {\bibfnamefont {C.}~\bibnamefont {Bannwarth}}, \
  and\ \bibinfo {author} {\bibfnamefont {P.}~\bibnamefont {Shushkov}},\
  }\href@noop {} {\bibfield  {journal} {\bibinfo  {journal} {Journal of
  chemical theory and computation}\ }\textbf {\bibinfo {volume} {13}},\
  \bibinfo {pages} {1989} (\bibinfo {year} {2017})}\BibitemShut {NoStop}%
\bibitem [{\citenamefont
  {Izutsu}(2009)}]{electrochemistry_in_nonaqueous_solutions}%
  \BibitemOpen
  \bibfield  {author} {\bibinfo {author} {\bibfnamefont {K.}~\bibnamefont
  {Izutsu}},\ }\href@noop {} {\emph {\bibinfo {title} {Electrochemistry in
  nonaqueous solutions}}}\ (\bibinfo  {publisher} {John Wiley \& Sons},\
  \bibinfo {year} {2009})\BibitemShut {NoStop}%
\bibitem [{\citenamefont {Inzelt}\ \emph {et~al.}(2013)\citenamefont {Inzelt},
  \citenamefont {Lewenstam},\ and\ \citenamefont
  {Scholz}}]{handbook_of_reference_electrodes}%
  \BibitemOpen
  \bibfield  {author} {\bibinfo {author} {\bibfnamefont {G.}~\bibnamefont
  {Inzelt}}, \bibinfo {author} {\bibfnamefont {A.}~\bibnamefont {Lewenstam}}, \
  and\ \bibinfo {author} {\bibfnamefont {F.}~\bibnamefont {Scholz}},\
  }\href@noop {} {\emph {\bibinfo {title} {Handbook of reference
  electrodes}}},\ Vol.\ \bibinfo {volume} {541}\ (\bibinfo  {publisher}
  {Springer},\ \bibinfo {year} {2013})\BibitemShut {NoStop}%
\bibitem [{\citenamefont {Chen}\ and\ \citenamefont
  {Guestrin}(2016)}]{XGBoost}%
  \BibitemOpen
  \bibfield  {author} {\bibinfo {author} {\bibfnamefont {T.}~\bibnamefont
  {Chen}}\ and\ \bibinfo {author} {\bibfnamefont {C.}~\bibnamefont
  {Guestrin}},\ }in\ \href@noop {} {\emph {\bibinfo {booktitle} {Proceedings of
  the 22nd acm sigkdd international conference on knowledge discovery and data
  mining}}}\ (\bibinfo {year} {2016})\ pp.\ \bibinfo {pages}
  {785--794}\BibitemShut {NoStop}%
\bibitem [{\citenamefont {Ramakrishnan}\ and\ \citenamefont {von
  Lilienfeld}(2017)}]{kernel_ridge_regression_in_quantum_chemistry}%
  \BibitemOpen
  \bibfield  {author} {\bibinfo {author} {\bibfnamefont {R.}~\bibnamefont
  {Ramakrishnan}}\ and\ \bibinfo {author} {\bibfnamefont {O.~A.}\ \bibnamefont
  {von Lilienfeld}},\ }\href@noop {} {\bibfield  {journal} {\bibinfo  {journal}
  {Reviews in computational chemistry}\ }\textbf {\bibinfo {volume} {30}},\
  \bibinfo {pages} {225} (\bibinfo {year} {2017})}\BibitemShut {NoStop}%
\bibitem [{\citenamefont {Huang}\ \emph {et~al.}(2018)\citenamefont {Huang},
  \citenamefont {Symonds},\ and\ \citenamefont {von
  Lilienfeld}}]{Fundamentals_of_quantum_machine_learning}%
  \BibitemOpen
  \bibfield  {author} {\bibinfo {author} {\bibfnamefont {B.}~\bibnamefont
  {Huang}}, \bibinfo {author} {\bibfnamefont {N.~O.}\ \bibnamefont {Symonds}},
  \ and\ \bibinfo {author} {\bibfnamefont {O.~A.}\ \bibnamefont {von
  Lilienfeld}},\ }\href@noop {} {\bibfield  {journal} {\bibinfo  {journal}
  {arXiv preprint arXiv:1807.04259}\ } (\bibinfo {year} {2018})}\BibitemShut
  {NoStop}%
\bibitem [{\citenamefont {Christensen}\ \emph {et~al.}(2017)\citenamefont
  {Christensen}, \citenamefont {Faber}, \citenamefont {Huang}, \citenamefont
  {Bratholm}, \citenamefont {Tkatchenko}, \citenamefont {Muller},\ and\
  \citenamefont {von Lilienfeld}}]{QML_library}%
  \BibitemOpen
  \bibfield  {author} {\bibinfo {author} {\bibfnamefont {A.~S.}\ \bibnamefont
  {Christensen}}, \bibinfo {author} {\bibfnamefont {F.}~\bibnamefont {Faber}},
  \bibinfo {author} {\bibfnamefont {B.}~\bibnamefont {Huang}}, \bibinfo
  {author} {\bibfnamefont {L.}~\bibnamefont {Bratholm}}, \bibinfo {author}
  {\bibfnamefont {A.}~\bibnamefont {Tkatchenko}}, \bibinfo {author}
  {\bibfnamefont {K.}~\bibnamefont {Muller}}, \ and\ \bibinfo {author}
  {\bibfnamefont {O.}~\bibnamefont {von Lilienfeld}},\ }\href@noop {}
  {\bibfield  {journal} {\bibinfo  {journal} {URL https://github.
  com/qmlcode/qml}\ } (\bibinfo {year} {2017})}\BibitemShut {NoStop}%
\bibitem [{skl()}]{sklearn_KRR}%
  \BibitemOpen
  \href@noop {} {\enquote {\bibinfo {title}
  {sklearn.kernel\_ridge.{K}ernel{R}idge --- scikit-learn.org},}\ }\bibinfo
  {howpublished}
  {\url{https://scikit-learn.org/stable/modules/generated/sklearn.kernel_ridge.KernelRidge.html}}\BibitemShut
  {NoStop}%
\bibitem [{\citenamefont {Bergstra}\ \emph {et~al.}(2013)\citenamefont
  {Bergstra}, \citenamefont {Yamins},\ and\ \citenamefont {Cox}}]{HYPEROPT}%
  \BibitemOpen
  \bibfield  {author} {\bibinfo {author} {\bibfnamefont {J.}~\bibnamefont
  {Bergstra}}, \bibinfo {author} {\bibfnamefont {D.}~\bibnamefont {Yamins}}, \
  and\ \bibinfo {author} {\bibfnamefont {D.}~\bibnamefont {Cox}},\ }in\
  \href@noop {} {\emph {\bibinfo {booktitle} {International conference on
  machine learning}}}\ (\bibinfo {organization} {PMLR},\ \bibinfo {year}
  {2013})\ pp.\ \bibinfo {pages} {115--123}\BibitemShut {NoStop}%
\bibitem [{\citenamefont {Behler}(2011)}]{atom_centered_symmetry_functions}%
  \BibitemOpen
  \bibfield  {author} {\bibinfo {author} {\bibfnamefont {J.}~\bibnamefont
  {Behler}},\ }\href@noop {} {\bibfield  {journal} {\bibinfo  {journal} {The
  Journal of chemical physics}\ }\textbf {\bibinfo {volume} {134}},\ \bibinfo
  {pages} {074106} (\bibinfo {year} {2011})}\BibitemShut {NoStop}%
\bibitem [{\citenamefont {Moriwaki}\ \emph {et~al.}(2018)\citenamefont
  {Moriwaki}, \citenamefont {Tian}, \citenamefont {Kawashita},\ and\
  \citenamefont {Takagi}}]{MORDRED}%
  \BibitemOpen
  \bibfield  {author} {\bibinfo {author} {\bibfnamefont {H.}~\bibnamefont
  {Moriwaki}}, \bibinfo {author} {\bibfnamefont {Y.-S.}\ \bibnamefont {Tian}},
  \bibinfo {author} {\bibfnamefont {N.}~\bibnamefont {Kawashita}}, \ and\
  \bibinfo {author} {\bibfnamefont {T.}~\bibnamefont {Takagi}},\ }\href@noop {}
  {\bibfield  {journal} {\bibinfo  {journal} {Journal of cheminformatics}\
  }\textbf {\bibinfo {volume} {10}},\ \bibinfo {pages} {1} (\bibinfo {year}
  {2018})}\BibitemShut {NoStop}%
\bibitem [{\citenamefont {Bart{\'o}k}\ \emph {et~al.}(2013)\citenamefont
  {Bart{\'o}k}, \citenamefont {Kondor},\ and\ \citenamefont
  {Cs{\'a}nyi}}]{smooth_overlap_of_atomic_positions}%
  \BibitemOpen
  \bibfield  {author} {\bibinfo {author} {\bibfnamefont {A.~P.}\ \bibnamefont
  {Bart{\'o}k}}, \bibinfo {author} {\bibfnamefont {R.}~\bibnamefont {Kondor}},
  \ and\ \bibinfo {author} {\bibfnamefont {G.}~\bibnamefont {Cs{\'a}nyi}},\
  }\href@noop {} {\bibfield  {journal} {\bibinfo  {journal} {Physical Review
  B}\ }\textbf {\bibinfo {volume} {87}},\ \bibinfo {pages} {184115} (\bibinfo
  {year} {2013})}\BibitemShut {NoStop}%
\bibitem [{\citenamefont {Huang}\ and\ \citenamefont {von
  Lilienfeld}(2020)}]{SLATM}%
  \BibitemOpen
  \bibfield  {author} {\bibinfo {author} {\bibfnamefont {B.}~\bibnamefont
  {Huang}}\ and\ \bibinfo {author} {\bibfnamefont {O.~A.}\ \bibnamefont {von
  Lilienfeld}},\ }\href@noop {} {\bibfield  {journal} {\bibinfo  {journal}
  {Nature chemistry}\ }\textbf {\bibinfo {volume} {12}},\ \bibinfo {pages}
  {945} (\bibinfo {year} {2020})}\BibitemShut {NoStop}%
\bibitem [{\citenamefont {Musil}\ \emph {et~al.}(2021)\citenamefont {Musil},
  \citenamefont {Grisafi}, \citenamefont {Bart{\'o}k}, \citenamefont {Ortner},
  \citenamefont {Cs{\'a}nyi},\ and\ \citenamefont
  {Ceriotti}}]{Physics_inspired_representations}%
  \BibitemOpen
  \bibfield  {author} {\bibinfo {author} {\bibfnamefont {F.}~\bibnamefont
  {Musil}}, \bibinfo {author} {\bibfnamefont {A.}~\bibnamefont {Grisafi}},
  \bibinfo {author} {\bibfnamefont {A.~P.}\ \bibnamefont {Bart{\'o}k}},
  \bibinfo {author} {\bibfnamefont {C.}~\bibnamefont {Ortner}}, \bibinfo
  {author} {\bibfnamefont {G.}~\bibnamefont {Cs{\'a}nyi}}, \ and\ \bibinfo
  {author} {\bibfnamefont {M.}~\bibnamefont {Ceriotti}},\ }\href@noop {}
  {\bibfield  {journal} {\bibinfo  {journal} {Chemical Reviews}\ }\textbf
  {\bibinfo {volume} {121}},\ \bibinfo {pages} {9759} (\bibinfo {year}
  {2021})}\BibitemShut {NoStop}%
\bibitem [{\citenamefont {Himanen}\ \emph {et~al.}(2020)\citenamefont
  {Himanen}, \citenamefont {J{\"a}ger}, \citenamefont {Morooka}, \citenamefont
  {Canova}, \citenamefont {Ranawat}, \citenamefont {Gao}, \citenamefont
  {Rinke},\ and\ \citenamefont {Foster}}]{DScribe}%
  \BibitemOpen
  \bibfield  {author} {\bibinfo {author} {\bibfnamefont {L.}~\bibnamefont
  {Himanen}}, \bibinfo {author} {\bibfnamefont {M.~O.}\ \bibnamefont
  {J{\"a}ger}}, \bibinfo {author} {\bibfnamefont {E.~V.}\ \bibnamefont
  {Morooka}}, \bibinfo {author} {\bibfnamefont {F.~F.}\ \bibnamefont {Canova}},
  \bibinfo {author} {\bibfnamefont {Y.~S.}\ \bibnamefont {Ranawat}}, \bibinfo
  {author} {\bibfnamefont {D.~Z.}\ \bibnamefont {Gao}}, \bibinfo {author}
  {\bibfnamefont {P.}~\bibnamefont {Rinke}}, \ and\ \bibinfo {author}
  {\bibfnamefont {A.~S.}\ \bibnamefont {Foster}},\ }\href@noop {} {\bibfield
  {journal} {\bibinfo  {journal} {Computer Physics Communications}\ }\textbf
  {\bibinfo {volume} {247}},\ \bibinfo {pages} {106949} (\bibinfo {year}
  {2020})}\BibitemShut {NoStop}%
\bibitem [{\citenamefont {Ruddigkeit}\ \emph {et~al.}(2012)\citenamefont
  {Ruddigkeit}, \citenamefont {Van~Deursen}, \citenamefont {Blum},\ and\
  \citenamefont {Reymond}}]{QM9_database2}%
  \BibitemOpen
  \bibfield  {author} {\bibinfo {author} {\bibfnamefont {L.}~\bibnamefont
  {Ruddigkeit}}, \bibinfo {author} {\bibfnamefont {R.}~\bibnamefont
  {Van~Deursen}}, \bibinfo {author} {\bibfnamefont {L.~C.}\ \bibnamefont
  {Blum}}, \ and\ \bibinfo {author} {\bibfnamefont {J.-L.}\ \bibnamefont
  {Reymond}},\ }\href@noop {} {\bibfield  {journal} {\bibinfo  {journal}
  {Journal of chemical information and modeling}\ }\textbf {\bibinfo {volume}
  {52}},\ \bibinfo {pages} {2864} (\bibinfo {year} {2012})}\BibitemShut
  {NoStop}%
\bibitem [{\citenamefont {Petersson}\ \emph {et~al.}(1988)\citenamefont
  {Petersson}, \citenamefont {Bennett}, \citenamefont {Tensfeldt},
  \citenamefont {Al-Laham}, \citenamefont {Shirley},\ and\ \citenamefont
  {Mantzaris}}]{Basis}%
  \BibitemOpen
  \bibfield  {author} {\bibinfo {author} {\bibfnamefont {a.}~\bibnamefont
  {Petersson}}, \bibinfo {author} {\bibfnamefont {A.}~\bibnamefont {Bennett}},
  \bibinfo {author} {\bibfnamefont {T.~G.}\ \bibnamefont {Tensfeldt}}, \bibinfo
  {author} {\bibfnamefont {M.~A.}\ \bibnamefont {Al-Laham}}, \bibinfo {author}
  {\bibfnamefont {W.~A.}\ \bibnamefont {Shirley}}, \ and\ \bibinfo {author}
  {\bibfnamefont {J.}~\bibnamefont {Mantzaris}},\ }\href@noop {} {\bibfield
  {journal} {\bibinfo  {journal} {The Journal of chemical physics}\ }\textbf
  {\bibinfo {volume} {89}},\ \bibinfo {pages} {2193} (\bibinfo {year}
  {1988})}\BibitemShut {NoStop}%
\bibitem [{\citenamefont {Lee}\ \emph {et~al.}(1988)\citenamefont {Lee},
  \citenamefont {Yang},\ and\ \citenamefont {Parr}}]{Functional}%
  \BibitemOpen
  \bibfield  {author} {\bibinfo {author} {\bibfnamefont {C.}~\bibnamefont
  {Lee}}, \bibinfo {author} {\bibfnamefont {W.}~\bibnamefont {Yang}}, \ and\
  \bibinfo {author} {\bibfnamefont {R.~G.}\ \bibnamefont {Parr}},\ }\href@noop
  {} {\bibfield  {journal} {\bibinfo  {journal} {Physical review B}\ }\textbf
  {\bibinfo {volume} {37}},\ \bibinfo {pages} {785} (\bibinfo {year}
  {1988})}\BibitemShut {NoStop}%
\bibitem [{\citenamefont {Becke}(1993)}]{Functional2}%
  \BibitemOpen
  \bibfield  {author} {\bibinfo {author} {\bibfnamefont {A.~D.}\ \bibnamefont
  {Becke}},\ }\href@noop {} {\bibfield  {journal} {\bibinfo  {journal} {The
  Journal of Chemical Physics. doi: https://doi. org/10.1063/1.464913}\ }
  (\bibinfo {year} {1993})}\BibitemShut {NoStop}%
\bibitem [{\citenamefont {Rajan}\ \emph {et~al.}(2020)\citenamefont {Rajan},
  \citenamefont {Zielesny},\ and\ \citenamefont {Steinbeck}}]{decimer}%
  \BibitemOpen
  \bibfield  {author} {\bibinfo {author} {\bibfnamefont {K.}~\bibnamefont
  {Rajan}}, \bibinfo {author} {\bibfnamefont {A.}~\bibnamefont {Zielesny}}, \
  and\ \bibinfo {author} {\bibfnamefont {C.}~\bibnamefont {Steinbeck}},\
  }\href@noop {} {\bibfield  {journal} {\bibinfo  {journal} {Journal of
  Cheminformatics}\ }\textbf {\bibinfo {volume} {12}},\ \bibinfo {pages} {1}
  (\bibinfo {year} {2020})}\BibitemShut {NoStop}%
\bibitem [{\citenamefont {Rajan}\ \emph
  {et~al.}(2021{\natexlab{a}})\citenamefont {Rajan}, \citenamefont {Zielesny},\
  and\ \citenamefont {Steinbeck}}]{decimer1}%
  \BibitemOpen
  \bibfield  {author} {\bibinfo {author} {\bibfnamefont {K.}~\bibnamefont
  {Rajan}}, \bibinfo {author} {\bibfnamefont {A.}~\bibnamefont {Zielesny}}, \
  and\ \bibinfo {author} {\bibfnamefont {C.}~\bibnamefont {Steinbeck}},\
  }\href@noop {} {\bibfield  {journal} {\bibinfo  {journal} {Journal of
  Cheminformatics}\ }\textbf {\bibinfo {volume} {13}},\ \bibinfo {pages} {1}
  (\bibinfo {year} {2021}{\natexlab{a}})}\BibitemShut {NoStop}%
\bibitem [{\citenamefont {Rajan}\ \emph
  {et~al.}(2021{\natexlab{b}})\citenamefont {Rajan}, \citenamefont {Brinkhaus},
  \citenamefont {Sorokina}, \citenamefont {Zielesny},\ and\ \citenamefont
  {Steinbeck}}]{decimer_segmentation}%
  \BibitemOpen
  \bibfield  {author} {\bibinfo {author} {\bibfnamefont {K.}~\bibnamefont
  {Rajan}}, \bibinfo {author} {\bibfnamefont {H.~O.}\ \bibnamefont
  {Brinkhaus}}, \bibinfo {author} {\bibfnamefont {M.}~\bibnamefont {Sorokina}},
  \bibinfo {author} {\bibfnamefont {A.}~\bibnamefont {Zielesny}}, \ and\
  \bibinfo {author} {\bibfnamefont {C.}~\bibnamefont {Steinbeck}},\ }\href@noop
  {} {\bibfield  {journal} {\bibinfo  {journal} {Journal of Cheminformatics}\
  }\textbf {\bibinfo {volume} {13}},\ \bibinfo {pages} {1} (\bibinfo {year}
  {2021}{\natexlab{b}})}\BibitemShut {NoStop}%
\bibitem [{\citenamefont {Amari}(1993)}]{Universal_theorem_on_learning_curves}%
  \BibitemOpen
  \bibfield  {author} {\bibinfo {author} {\bibfnamefont {S.-I.}\ \bibnamefont
  {Amari}},\ }\href@noop {} {\bibfield  {journal} {\bibinfo  {journal} {Neural
  networks}\ }\textbf {\bibinfo {volume} {6}},\ \bibinfo {pages} {161}
  (\bibinfo {year} {1993})}\BibitemShut {NoStop}%
\bibitem [{\citenamefont {Miller}\ \emph {et~al.}(1972)\citenamefont {Miller},
  \citenamefont {Nordblom},\ and\ \citenamefont {Mayeda}}]{Ref_05}%
  \BibitemOpen
  \bibfield  {author} {\bibinfo {author} {\bibfnamefont {L.~L.}\ \bibnamefont
  {Miller}}, \bibinfo {author} {\bibfnamefont {G.}~\bibnamefont {Nordblom}}, \
  and\ \bibinfo {author} {\bibfnamefont {E.~A.}\ \bibnamefont {Mayeda}},\
  }\href@noop {} {\bibfield  {journal} {\bibinfo  {journal} {The Journal of
  Organic Chemistry}\ }\textbf {\bibinfo {volume} {37}},\ \bibinfo {pages}
  {916} (\bibinfo {year} {1972})}\BibitemShut {NoStop}%
\bibitem [{\citenamefont {Weinberg}\ and\ \citenamefont
  {Weinberg}(1968)}]{Ref_24}%
  \BibitemOpen
  \bibfield  {author} {\bibinfo {author} {\bibfnamefont {N.}~\bibnamefont
  {Weinberg}}\ and\ \bibinfo {author} {\bibfnamefont {H.}~\bibnamefont
  {Weinberg}},\ }\href@noop {} {\bibfield  {journal} {\bibinfo  {journal}
  {Chemical Reviews}\ }\textbf {\bibinfo {volume} {68}},\ \bibinfo {pages}
  {449} (\bibinfo {year} {1968})}\BibitemShut {NoStop}%
\bibitem [{\citenamefont {Mazouin}\ \emph {et~al.}(2022)\citenamefont
  {Mazouin}, \citenamefont {Sch{\"o}pfer},\ and\ \citenamefont {von
  Lilienfeld}}]{HOMO_LUMO_gap_energies_QM9}%
  \BibitemOpen
  \bibfield  {author} {\bibinfo {author} {\bibfnamefont {B.}~\bibnamefont
  {Mazouin}}, \bibinfo {author} {\bibfnamefont {A.~A.}\ \bibnamefont
  {Sch{\"o}pfer}}, \ and\ \bibinfo {author} {\bibfnamefont {O.~A.}\
  \bibnamefont {von Lilienfeld}},\ }\href@noop {} {\bibfield  {journal}
  {\bibinfo  {journal} {Materials Advances}\ }\textbf {\bibinfo {volume} {3}},\
  \bibinfo {pages} {8306} (\bibinfo {year} {2022})}\BibitemShut {NoStop}%
\bibitem [{\citenamefont {Abolhasani}\ and\ \citenamefont
  {Kumacheva}(2023)}]{self_driving_labs}%
  \BibitemOpen
  \bibfield  {author} {\bibinfo {author} {\bibfnamefont {M.}~\bibnamefont
  {Abolhasani}}\ and\ \bibinfo {author} {\bibfnamefont {E.}~\bibnamefont
  {Kumacheva}},\ }\href@noop {} {\bibfield  {journal} {\bibinfo  {journal}
  {Nature Synthesis}\ ,\ \bibinfo {pages} {1}} (\bibinfo {year}
  {2023})}\BibitemShut {NoStop}%
\end{thebibliography}%

\end{document}